\documentclass[10pt,twocolumn]{article}

\usepackage[a4paper,left=18mm,right=18mm,top=22mm,bottom=28mm]{geometry}

\usepackage{times}
\usepackage{microtype}
\usepackage{amsmath}
\usepackage{graphicx}
\usepackage{float}
\usepackage{enumitem}
\usepackage{multirow}
\usepackage{colortbl}

\usepackage{tabularx}
\usepackage{array}
\newcolumntype{Y}{>{\raggedright\arraybackslash}X}

\usepackage[numbers,sort&compress]{natbib}
\usepackage{fancyhdr}
\AtBeginDocument{%
  }

\usepackage{xcolor}
\newcommand{\revision}[1]{#1} 

\usepackage[T1]{fontenc}
\usepackage[hyphens]{url} 
\newcommand{\doi}[1]{doi:\,\url{https://doi.org/#1}}

\usepackage[hidelinks]{hyperref}

\begin{document}

\title{Hacking Flow: From Lived Practices to Innovation}

\author{
Fabio Stano$^{1}$ \and
Max L Wilson$^{2}$ \and
Christof Weinhardt$^{1}$ \and
Michael T Knierim$^{1,2}$
}

\date{
$^{1}$ Institute for Information Systems (WIN), Karlsruhe Institute of Technology (KIT), Germany\\
$^{2}$ School of Computer Science, University of Nottingham, United Kingdom\\
\texttt{fabio.stano@kit.edu}\\[0.5em]
\textbf{Preprint.}\\
Accepted for publication at CHI 2026. The final version will appear in the ACM Digital Library.
}

\maketitle

\begin{abstract}
In digital knowledge work, flow promises not just productivity; it offers a pathway to well-being. Yet despite decades of flow research in HCI, we know little about how to design digital interventions that support it. In this work, we foreground lived interventions — everyday practices workers already use to foster flow — to uncover overlooked opportunities and chart new directions for digital intervention design. Specifically, we report findings from two studies: (1) a reflexive thematic analysis of open-ended survey responses (n = 160), surfacing 38 lived interventions across four categories: environment, organization, task shaping, and personal readiness; and (2) a quantitative online survey (n = 121) that validates this repertoire, identifies which interventions are broadly endorsed versus polarizing, and elicits visions of technological support. We contribute empirical insights into how digital workers cultivate flow, situate these lived interventions within existing literature, and derive design opportunities for future digital flow interventions.
\end{abstract}

\section{Introduction}

In today’s digitally mediated work environments, employees face increased cognitive demands due to complex and frequent information exchange, interruptions, and reduced socio-affective support \cite{bartholomeyczikFosteringFlowExperiences2023a, markCostInterruptedWork2008, taserExaminationRemoteEworking2022}. In turn, fostering sustained focus, well-being, and motivation has become a critical concern for organizations and individuals \cite{bartholomeyczikFosteringFlowExperiences2023a}. One promising approach to addressing these challenges is through cultivating flow, a state of deep immersion and intrinsic motivation \cite{csikszentmihalyiFlowPsychologyOptimal1990, nadjWhatDisruptsFlow2023}. While traditionally examined in leisure contexts, recent work highlights its frequent occurrence in professional settings \cite{brownUsingLogsData2023, cowleySeekingFlowFinegrained2022, pastushenkoMethodologyMultimodalLearning2020, yotsidiAddFlowFire2018, aferganDynamicDifficultyUsing2014, zugerReducingInterruptionsWork2017a}. In the workplace, flow offers a unique promise: it combines productivity with positive experience, enabling knowledge workers not only to perform efficiently but also to feel energized, satisfied, and protected against burnout \cite{bartholomeyczikFosteringFlowExperiences2023a}.

In HCI, flow has served as a lens to evaluate user experiences and as a design goal to optimize engagement across domains such as gaming, education, creativity, and e-commerce \cite{websterDimensionalityCorrelatesFlow1993, paceGroundedTheoryFlow2004, weibelPlayingOnlineGames2008, hamariChallengingGamesHelp2016, oliveiraAutomaticFlowExperience2019, skadbergVisitorsFlowExperience2004}. More recently, research has made strides in flow detection, for example through log-based indicators of attention or physiological sensing with heart rate and EEG \cite{brownUsingLogsData2023, risslerGotFlowUsing2018, knierimExploringFlowRealWorld2025a}. These advances enable systems that monitor user states and adjust conditions in real time. Yet when it comes to interventions (i.e., how to actively foster and maintain flow at work), the design space remains largely underexplored. In our \revision{review of 30 years of flow research in major HCI outlets}, out of 96 articles, only 5 implemented and tested a digital intervention designed to assist workplace flow. Existing systems provide valuable point solutions (e.g., FlowLight \cite{zugerReducingInterruptionsWork2017a}), but exploration of the broader intervention space is still missing.

However, innovation may already be hiding in plain sight. Prior HCI research on situated practice and appropriation shows that people routinely bend, repurpose, and domesticate technologies; work on folk theories and lived informatics documents how everyday regulation, tracking, and self-management become woven into daily well-being routines \cite{suchmanHumanMachineReconfigurationsPlans2006, crabtreeDoingDesignEthnography2012, eslamiFirstItThen2016, devitoHowPeopleForm2018, rooksbyPersonalTrackingLived2014}. Applying this perspective, knowledge workers are not passive recipients of technology but active agents who experiment with and sustain practices to regulate attention and effort in digital contexts. We call these \emph{lived interventions}: practices, tactics, or adjustments that workers have devised and integrated into their everyday routines to foster flow. Recognizing workers as domain experts in their own contexts motivates studying these lived interventions systematically, and examining how individual differences in flow metacognition \cite{wilsonFlowMetacognitionsQuestionnaire2016a} relate to their endorsement.

Our research goal is to identify a repertoire of lived interventions that digital workers use. To guide this effort, we ask two research questions:  
\begin{enumerate}
    \item[] \textbf{RQ1:} How do digital knowledge workers foster flow through lived interventions, and how are these interventions disseminated, endorsed, or disendorsed across workers?  
    \item[] \textbf{RQ2:} Which of these lived interventions have already been addressed in existing research and digital intervention design, and where do uncovered opportunities for technological support remain?  
\end{enumerate}

We report findings from two complementary studies in a sequential exploratory design \cite{creswellAdvancedMixedMethods2003}. The first is a reflexive thematic analysis of open-ended survey responses (n=80+80) that surfaces a diverse repertoire of 38 lived interventions across four themes: optimizing the environment, shaping the task, planning and organization, and ensuring mental and physical readiness. The second study is a survey (n=121) that uses a crowd-evaluation approach to quantitatively identify consensus and conflict in strategy endorsement, and to examine how workers envision technological support for them. We further contrast those results to the state of the art of flow intervention research, highlighting both promising directions and uncovered gaps. Altogether, our work contributes:
\begin{enumerate}
    \item Empirical insights into the diverse, situated ways in which digital workers actively foster flow.
    \item Design opportunities that emerge when mapping lived interventions against existing work and participants’ visions, showing where future systems could extend beyond current solutions.
\end{enumerate}

By combining empirical grounding with actionable design directions, we aim to advance and speak to HCI’s unique ability to design systems that not only protect workers from interruptions, but actively help them to foster flow in digital work across multiple dimensions.
\section{Theoretical Foundations and Related Work}

\subsection{Everyday practices, appropriation, folk theories, and lived informatics}
HCI has a long tradition of examining the everyday practices people devise to make technology and work fit their lives. Early ethnographic and workplace studies foregrounded situated practices, meaning the ways actual work departs from formal procedures and plans \cite{suchmanHumanMachineReconfigurationsPlans2006, crabtreeDoingDesignEthnography2012}. Building on this, HCI work on appropriation shows how users bend and repurpose technologies and outlines how to design for emergent use \cite{dixDesigningAppropriation2007, tchounikineDesigningAppropriationTheoretical2017}. In parallel, scholars have examined folk theories (users’ lay explanations for how socio-technical systems operate), which then guide strategy making \cite{eslamiFirstItThen2016, devitoHowPeopleForm2018}. More recently, the lived informatics perspective shows how tracking, self-management, and everyday regulation become woven into ongoing life, with practices evolving alongside constraints, resources, and identities \cite{rooksbyPersonalTrackingLived2014}. Across these strands, users are treated as agents who devise, adapt, and sustain practices to regulate attention, effort, and well-being in technology-mediated contexts. This is especially salient in digital knowledge work, where people must manage interruptions, motivation, and task difficulty (see also \citet{klasnjaHowEvaluateTechnologies2011} on evaluating behavior-change technologies). \revision{Taken together, these perspectives underscore the value of attending to people’s lived experiences, as they reveal the concrete, situated practices through which individuals regulate their work and well-being and offer actionable insights for design. This is an aspect that prior flow-at-work research has largely overlooked.}

\subsection{Flow Theory}
\begin{figure*}[t]
    \centering
    \includegraphics[width=0.9\linewidth]{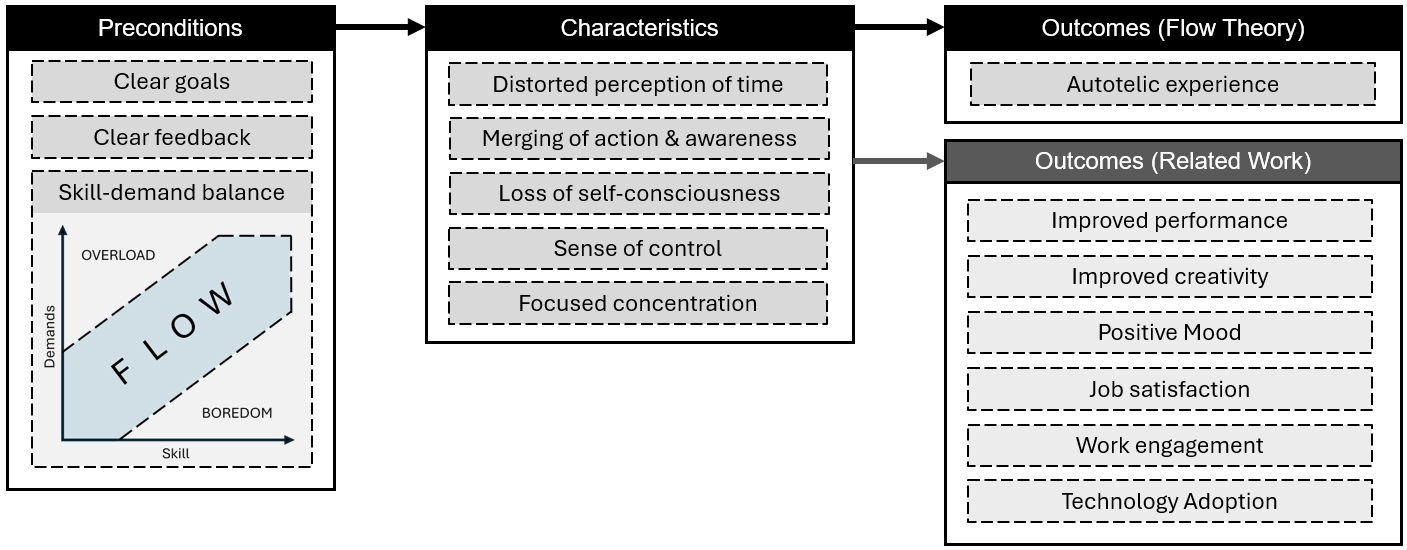}
    \caption{Theoretical Preconditions, Characteristics, and Outcomes of Flow}
    \label{fig:Flow Theory}
\end{figure*}

First introduced by \citet{csikszentmihalyiBoredomAnxiety1975}, flow refers to a state of complete immersion and total involvement in an activity. Over the decades, this concept has been extensively examined across various disciplines, forming a strong theoretical basis. Flow theory organizes the phenomenon into nine dimensions, grouped into preconditions, experiential state characteristics, and resulting outcomes (\cite{peiferAdvancesFlowResearch2021a} – see Figure~\ref{fig:Flow Theory}). Among these dimensions, the balance between challenge and skill has drawn considerable attention. This balance not only serves to map experiences - ranging from boredom (low challenge) to anxiety (high challenge) - but also provides a framework for adjusting and intensifying individual flow states (\cite{peiferAdvancesFlowResearch2021a} – see Figure~\ref{fig:Flow Theory}). In addition to this balance, clear goals provide a sense of direction and purpose, enabling individuals to focus their efforts effectively \cite{csikszentmihalyiFlowPsychologyOptimal1990}. Similarly, immediate and unambiguous feedback helps to sustain engagement by reducing uncertainty and allowing individuals to adjust their actions in real time \cite{csikszentmihalyiFlowPsychologyOptimal1990}.
The benefits of experiencing flow are well-documented and center around its autotelic nature; an inherently rewarding experience in which the activity itself becomes the primary source of enjoyment and fulfillment \cite{csikszentmihalyiBoredomAnxiety1975}. This state typically leads to increased intrinsic motivation, greater satisfaction, and enhanced performance \cite{csikszentmihalyiFlowPsychologyOptimal1990, engeserFlowPerformanceModerators2008, landhausserFlowItsAffective2012}. While often associated with leisure activities, studies suggest that flow occurs most frequently during work tasks, especially those involving complex problem-solving or creative demands \cite{engeserFluctuationFlowAffect2016a, csikszentmihalyiOptimalExperienceWork1989, quinnFlowKnowledgeWork2005}. These cognitive activities inherently meet the prerequisites for flow, making work settings a particularly valuable domain for studying the phenomenon. Professionals who regularly experience flow in their roles often report greater satisfaction \cite{maeranFlowExperienceJob2013}, higher energy levels \cite{demeroutiWorkrelatedFlowEnergy2012} and positive mood \cite{fullagarFlowWorkExperience2009}. Flow has also been linked to increased technology adoption \cite{dengUserExperienceSatisfaction2010, koufarisApplyingTechnologyAcceptance2002, lowryProposingHedonicAffect2013},  higher levels of task performance \cite{mahnkeFlowExperienceInformation2014, nadjWhatDisruptsFlow2023} and greater satisfaction in technology use \cite{cipressoPsychometricModelingPervasive2015, dengUserExperienceSatisfaction2010}. 
\revision{Beyond these benefits, recent work has started to examine people's metacognitive knowledge and beliefs about flow as a state of optimal functioning. Wilson and Moneta conceptualize such “flow metacognitions” as beliefs that being in flow fosters achievement and confidence in one's ability to self-regulate flow, and show that these beliefs can be measured reliably and predict the intensity and frequency of flow at work above and beyond general metacognitions and dispositional flow \cite{wilsonFlowMetacognitionsQuestionnaire2016a}.
}

\subsection{Flow in HCI}
\revision{To outline prior work on flow in HCI, we conducted a literature review, searching the ACM Digital Library (CSS: Human-Computer-Interaction), and the digital libraries of the International Journal of Human-Computer Studies, International Journal of Human–Computer Interaction, and Computers in Human Behavior, using the terms “flow state” OR “flow experience” in the abstract. We retained only articles that treated flow as a central variable (n = 96). Of these, 28 addressed workplace contexts, and only 5 described systems or interventions explicitly designed to foster flow at work. In the others,} HCI has examined flow both as an explanatory lens and as a design goal in other contexts. 
Much of the extant work has concentrated on applied domains, most prominently gaming (e.g., \cite{weibelPlayingOnlineGames2008, procciMeasuringFlowExperience2012}), education (e.g., \cite{giannakosFitbitLearningCapturing2020, choiERPTrainingWebbased2007}), and their intersection in game-based learning (e.g., \cite{admiraalConceptFlowCollaborative2011, hamariChallengingGamesHelp2016}). In gaming, studies show that competitive dynamics and social presence can enhance the intensity of flow \cite{weibelPlayingOnlineGames2008}. In education, large-scale surveys demonstrate that flow predicts not only enjoyment but also persistence and academic performance \cite{hamariChallengingGamesHelp2016}. In gamification, flow has been established as a key mechanism through which game-like elements drive motivation and learning outcomes \cite{hamariMeasuringFlowGamification2014, uhmStimulatingSuspenseGamified2023, liewGameChangerNPCsLevelingUp2024, kloepGamificationManufacturingWork2025, chenHowGamificationSerendipity2025}. Beyond these domains, flow has also been studied in contexts such as commerce \cite{skadbergVisitorsFlowExperience2004, liuEnhancingFlowExperience2016}, social media \cite{brailovskaiaItNeedIt2020, kaurAssessingFlowExperience2016}, and immersive technologies \cite{dreySpARklingPaperEnhancingCommon2022, kimImpactVirtualReality2019, bianExploringWeakAssociation2018}. 

More recently, HCI research has shifted toward making flow computationally accessible to enable adaptive systems. Scholars have explored log data of user actions \cite{brownUsingLogsData2023}, physiological signals such as ECG and heart-rate variability \cite{risslerGotFlowUsing2018}, and neural measures using EEG \cite{labonte-lemoyneAreWeFlow2016, knierimExploringFlowRealWorld2025a}. This work moves beyond treating flow solely as a design outcome toward operationalizing it as a real-time input for system adaptation \cite{nadjWhatDisruptsFlow2023, zugerInterruptibilitySoftwareDevelopers2015a}. Despite this growing body of work, the design of systems that deliberately foster flow has remained comparatively rare.

In parallel, workplace studies increasingly recognize flow as a central aspect of digital labor. For example, Taser et al. \cite{taserExaminationRemoteEworking2022} found that while remote work can elevate flow, stressors such as technostress and loneliness can diminish workers’ ability to concentrate. A recurring theme in workplace research concerns interruptions and their disruptive effect on flow (e.g. \cite{ritonummiFlowExperienceSoftware2023}). Züger et al. \cite{zugerReducingInterruptionsWork2017a} showed how developers’ ability to re-enter flow after task switching can be supported by recovery cues, while physiological work further documented how workplace notifications and interruptions are reflected in heart rate variability and EEG measures of flow disruption \cite{risslerGotFlowUsing2018}. Also, immersive technologies have been studied as tools to cultivate flow at work. Experiments with virtual and augmented reality show that immersive environments not only increase presence but also sustain higher levels of concentration and absorption during knowledge-intensive tasks \cite{kimImpactVirtualReality2019}. Relatedly, gamification of work has been researched, with recent work \cite{kloepGamificationManufacturingWork2025} demonstrating that game-like elements can also boost challenge–skill alignment at work and positively affect employees’ flow experiences.

\subsection{Flow Interventions}
Recent research highlights that while the conditions for flow (i.e., clear goals, immediate feedback, and a balance between skills and challenges) are well established, translating this knowledge into practical interventions requires further exploration \cite{peiferAdvancesFlowResearch2021a, bartholomeyczikFosteringFlowExperiences2023a}. Based on theory, these preconditions naturally suggest themselves as intervention targets. In particular, manipulating task difficulty has been at the forefront of adaptive technology research, as it offers a salient and conceptually simple way of aligning challenge and skill (e.g., \cite{chanelEmotionAssessmentPhysiological2011, aferganDynamicDifficultyUsing2014, ewingEvaluationAdaptiveGame2016}. 
To systematize the development of interventions in professional contexts, Bartholomeyczik et al. \cite{bartholomeyczikFosteringFlowExperiences2023a} proposed a structured framework. After reviewing existing work, they call for more (in particularly adaptive) flow interventions tailored to the realities of work \cite{bartholomeyczikFosteringFlowExperiences2023a}.

So far, flow interventions have been most studied in domains like sports and exercise, where they are designed to enhance performance and positive experiences. For example, Goddard et al. \cite{goddardSystematicReviewFlow2023a} identified and systematically reviewed 29 studies employing strategies such as mindfulness, hypnosis, and imagery to promote flow. These methods are often hardly applicable to a digital work context, due to the different nature of the tasks, where a lot of physical movement is involved.
In contrast, the body of research on flow interventions at work remains thinner and is predominantly grounded in psychology. SMART goal-setting \cite{weintraubNudgingFlowSMART2021a} and job-crafting interventions \cite{costantiniImplementingJobCrafting2022a} have been shown to increase flow alongside engagement and performance, while digital positive-psychology programs \cite{drozdBetterDaysRandomized2014a} and humor-based exercises in nursing \cite{bartzikCareJoyEvaluation2021a} likewise reported flow as a beneficial outcome. Together, these studies demonstrate both the promise and the contextual challenges of designing effective flow interventions for knowledge work. 
Within Information Systems research, a small but growing strand explores technology-driven interventions. For example, Adam et al. \cite{adamHumanCenteredDesignEvaluation2024a} developed neuroadaptive systems that leverage physiological measures to manage interruptions, embedding flow theory directly into IT design. This illustrates how adaptive, technology-based interventions can move beyond psychological training toward system-level support, providing a gateway to the HCI-specific approaches discussed in the next subsection.

\subsection{Computer Assisted Flow Interventions}

Most existing technology-based flow interventions have centered on reducing workplace distractions. \revision{For example,} Züger et al. \cite{zugerReducingInterruptionsWork2017a} introduced FlowLight, a system that uses physical LED indicators and automatic interruptibility measures to reduce workplace interruptions. Their large-scale study demonstrated a 46\% reduction in interruptions, highlighting the system’s effectiveness in creating conditions conducive to flow. Kim et al. \cite{kimTechnologySupportedBehavior2017} developed a behavior-restriction system that helps users block digital distractions, showing that actively limiting potentially disruptive technologies can strengthen conditions for flow. Similarly, Ruvimova et al. \cite{ruvimovaTransportMeAway2020} demonstrated that immersive VR environments can mitigate the distractions of open offices, enabling knowledge workers to maintain focus and experience higher levels of flow despite noisy surroundings. Relatedly, Cuber et al. \cite{cuberExaminingUseVR2024} examined VR as a study aid for university students with ADHD, reporting sustained increases in concentration, motivation, and effort. While conducted in an academic context, the tasks and demands closely mirror those of digital knowledge work, suggesting strong transferability of findings.

More recent work illustrates that flow support in HCI need not be limited to shielding users from interruptions. Bartholomeyczik et al. \cite{bartholomeyczikFosteringFlowExperiences2023a} developed and examined a smartphone-based, just-in-time adaptive intervention to foster flow among knowledge workers. The intervention prompted participants to use metacognitive strategies, such as mental contrasting or simple goal-setting, to align their tasks with their skills. While both strategies increased the likelihood of experiencing flow, mental contrasting showed greater effectiveness over time. Beyrodt et al. \cite{beyrodtSociallyInteractiveAgents2023} evaluated socially interactive cobot avatars in manufacturing work, showing that collaborative digital agents can scaffold workers’ focus and engagement. While situated in physical production settings, their findings highlight design principles that could be transferred to digital workplaces where repetitive or monotonous tasks also threaten flow. Complementing this, Kloep et al. \cite{kloepGamificationManufacturingWork2025} demonstrated that gamification elements in manufacturing tasks increased the frequency of flow experiences. Together, these studies highlight both the narrow focus of much existing work on distraction reduction and the untapped potential of alternative approaches to flow-supportive system design.

\section{Study 1: Building a Repertoire of Lived Flow Interventions}
Study 1 asked a simple but central question: \emph{What do people do to foster flow in their daily work?} To answer this, we collected open-ended survey responses about participants’ strategies, habits, and tools, and analyzed them thematically to surface a structured repertoire of interventions.

\subsection{Method}
\subsubsection{Study Design}
The study was implemented as an online survey in oTree \cite{chenOTreeOpensourcePlatform2016a}, to gather insights into the practices of a broad pool of people. After giving \revision{informed} consent, participants were given an established definition of the Flow Experience (see \cite{wilsonFlowMetacognitionsQuestionnaire2016a}), to make sure they understand the concept. Then they were asked to keep that definition in mind for the rest of the study, and think about how they experience it at work. After that, participants were asked to \revision{provide basic demographic information\footnote{Participants recruited via Prolific did not complete these items because Prolific automatically supplies demographic data.} and} fill out the Short Flow in Work Scale (SFWS; \citet{monetaValidationShortFlow2017a}), to measure the occurrence and intensity of Flow Experiences during their everyday work,and the Flow Metacognitions Questionnaire (FMQ; \citet{wilsonFlowMetacognitionsQuestionnaire2016a}), which measures both positive beliefs about flow (e.g., its benefits for productivity and creativity) and perceived control beliefs (e.g., the ability to enter or sustain flow). The questionnaires are presented in Table~\ref{tab:questionnaires}.

After this introduction to the flow experience, participants responded to the open-ended question: 

\begin{quote}
\emph{``What do you do to foster flow experiences in your daily work?''}
\end{quote}

This question was designed to elicit rich, experience-based responses focused on intentional practices that enhance flow. It was the main subject of study for the thematic analysis.

\begin{table*}[t]
\centering
\caption{Items of the Flow Metacognitions Questionnaire (FMQ \cite{wilsonFlowMetacognitionsQuestionnaire2016a}) and Short Flow in Work Scale (SFWS \cite{monetaValidationShortFlow2017a}).}
\label{tab:questionnaires}

\begin{tabular*}{\textwidth}{@{\extracolsep{\fill}} p{0.9cm} p{0.6cm} p{0.70\textwidth} p{2.8cm}}
\hline
\textbf{Scale} & \textbf{Item} & \textbf{Wording} & \textbf{Measurement} \\
\hline

FMQ & 1  & I become completely focused on the task when I am in flow.
& \multirow{12}{2.8cm}{4-point Likert (do not agree -- agree very much)} \\
    & 2  & I know how I can re-create having flow if I want to. & \\
    & 3  & Flow has a positive effect on the activity. & \\
    & 4  & I am able to quickly re-enter flow if I need to. & \\
    & 5  & I am able to generate various ideas and options while being in flow. & \\
    & 6  & Once I start with the activity there is no stopping me getting into flow. & \\
    & 7  & I know that by being in flow I achieve more. & \\
    & 8  & I know what I need to do to get into flow. & \\
    & 9  & My thinking becomes clearer when I am in flow. & \\
    & 10 & It is in my power to control when I have flow. & \\
    & 11 & I am more creative when I am in flow. & \\
    & 12 & I am able to sustain flow for long periods. & \\
\hline

SFWS & 1 & When I get really involved in my work my concentration becomes like my breathing \dots\ I never think of it.
& \multirow{3}{2.8cm}{4-point Likert (never/almost never true -- always/almost always true)} \\
     & 2 & Sometimes when I am working I become so absorbed that I am less aware of myself and my problems. & \\
     & 3 & When I am working I am so involved in it that I don’t see myself as separate from what I am doing. & \\
\hline
\end{tabular*}
\end{table*}

\subsubsection{Analytic Approach}
Based on the responses of Sample~1, we conducted a reflexive thematic analysis following the six-phase model outlined by Braun and Clarke \cite{braun2013successful}, adopting a contextualist position that acknowledges both the reality of participants’ experiences and the influence of social and psychological contexts on how these experiences are constructed. The analysis was primarily inductive and semantic in nature, focusing on participants’ expressed practices and interpretations without applying a predefined theoretical lens or coding framework.

\revision{Both analysts have developed substantial expertise in flow and digital knowledge work, reflected in several years of published research on these topics.
This background informed our familiarity with the practices and challenges participants described and shaped the analytic sensitivity we brought to the data. Working from a contextualist perspective, we acknowledge that our prior research on flow and work processes influenced the questions we asked of the material and the meanings we attended to. Throughout the analysis, we kept awareness of these assumptions and ensured that the resulting themes remained grounded in participants’ accounts.}

\begin{enumerate}
    \item[(1)] \textbf{Familiarization.} Two researchers independently read and re-read the entire dataset to immerse themselves in the data and begin identifying patterns of meaning.
    
    \item[(2)] \textbf{Generating Initial Codes.} Initial codes were generated manually by the researchers to capture meaningful units related to participants’ strategies for inducing flow. 31 codes were developed, reflecting concrete flow-supporting practices such as 'reducing distractions,' 'setting time limits,' or 'using music.' Where applicable, codes were enriched with examples of tools and technologies participants mentioned as supports (e.g., Pomodoro timers, Do-Not-Disturb mode, focus music playlists).
    
    \item[\revision{(3-5)}] \textbf{Searching, Reviewing, and Defining Themes.} Codes were iteratively grouped into four higher-order themes, reflecting broader categories of intervention. Themes were refined collaboratively through discussions between the coders, with careful attention to internal coherence and distinction across themes. Reflexive discussions were held throughout to surface and examine assumptions, ensuring the themes remained grounded in the data. 
    
    \item[(6)] \textbf{Producing the Report.} The final thematic structure was reviewed against the entire dataset to ensure representativeness and analytic depth. The identified themes served as the foundation for subsequent quantitative and design-oriented studies.
\end{enumerate}

Because Sample~1 was limited to students, we collected Sample~2 to validate the thematic structure in a more diverse working population. The prior codebook was used deductively to code the new data, with openness to new codes. The existing structure proved robust, with most responses fitting well into the previously established codes and themes. Only a few additional codes were derived, each with low frequency, further supporting the transferability of the initial thematic framework, while enriching it with more fine-grained information. The new codes also fit seamlessly into the existing themes.

\revision{We call the codes “lived interventions”, referring to the concrete, experience-based actions that participants described as helping them sustain or recover flow. The term emphasizes that the strategies are lived, since they arise from situated experience rather than from formal prescriptions, and that they function as interventions, since workers apply them deliberately to regulate their flow state. While the notion overlaps with established HCI concepts such as everyday and situated practices \cite{suchmanHumanMachineReconfigurationsPlans2006}, appropriation \cite{dixDesigningAppropriation2007, tchounikineDesigningAppropriationTheoretical2017}, folk theories \cite{eslamiFirstItThen2016, devitoHowPeopleForm2018}, and lived informatics \cite{ rooksbyPersonalTrackingLived2014}, our usage provides a clear label for presenting the themes of our analysis and structuring the resulting design space. We treat lived interventions not as a theoretical construct (like e.g. flow antecedents \cite{csikszentmihalyiFlowPsychologyOptimal1990} or metacognitive regulation \cite{dinsmoreFocusingConceptualLens2008}, which describe conceptual preconditions or cognitive processes), but as the concrete actions through which workers enact these processes in everyday practice.}

\subsubsection{Participants}
Participants were recruited through the student panel of a European university (Sample 1, n=80) and worldwide via Prolific (Sample 2, n=97), with both studies including only participants who self-identified as at least sometimes experiencing flow at work\footnote{\revision{The study invitation explicitly stated that we were interested in workers who sometimes experience flow. At the beginning of the survey, participants were additionally asked how often they experienced flow at work in the past three months. Individuals who responded “never” were thanked for their time and could end the study immediately while still receiving full compensation.}}. \revision{They were compensated with €2 or £1 for their participation, respectively.} Sample 1 consists of university students with regular experience of flow in academic or job-related work. Sample 2 consists of a more diverse general population, recruited through Prolific. We screened individuals who work at least 75\% of their time on a computer. \revision{This reflects the reality that modern knowledge work is predominantly mediated by digital tools and also corresponds to the closest available screening option on Prolific for identifying participants whose primary job activities take place through digital technologies.} Detailed sample statistics are displayed in Table~\ref{tab:demographics}. From the prolific sample, we had to exclude 17 participants due to obvious LLM responses. These were either characterized as still including the direct answer to the user's prompt or as comprising almost the exact same list of very specific interventions.

\begin{table*}[t]
\centering
\caption{Demographic characteristics of participants in Sample~1 and Sample~2.}
\label{tab:demographics}

\setlength{\tabcolsep}{8pt}

\begin{tabular}{l rr rrr rrr}
\hline
 & \multicolumn{2}{c}{\textbf{Sex}} 
 & \multicolumn{3}{c}{\textbf{Age (years)}} 
 & \multicolumn{3}{c}{\textbf{Employment status}} \\
\cline{2-3} \cline{4-6} \cline{7-9}
\textbf{} & Female & Male & Mean & Min & Max & Full-time & Part-time & Unemployed \\
\hline
\textbf{Sample 1} & 22 & 46 & 23.81 & 19 & 45 & 1  & 46 & 33 \\
\textbf{Sample 2} & 48 & 49 & 33.22 & 19 & 64 & 79 & 18 & 0  \\
\hline
\textbf{Total}    & 70 & 95 & 29 & 19 & 64 & 80 & 64 & 33 \\
\hline
\end{tabular}
\end{table*}


\subsection{Results}
\revision{Our analysis surfaced a wide variety of lived interventions, which participants described to foster flow in their everyday digital work. We report raw code frequencies rather than percentages, as our main aim was not to determine prevalence but to capture the breadth of lived interventions.} Because participants freely associated strategies from their individual experience, the data reflect high personal variation, making raw counts more informative than relative percentages, which would naturally be quite low. 

Study 1 yielded a total of 38 distinct codes contributed by 160 participants. On average, participants mentioned 2.47 interventions (Median = 2; range = 1-8). 71 participants reported within one category, 50 participants across two categories, 26 participants across three, and two participants reported interventions across all four categories. The four categories distinguish whether interventions are primarily focused on the environment, the temporal and organizational aspects of work, the task at hand, or the physical and mental readiness of the person themselves. In the following paragraphs, we describe each category in turn, outlining its range of interventions, relative prominence in the data, and illustrative examples. 

\subsubsection{Optimizing the Environment}

\begin{figure}[H]
    \centering
    \includegraphics[width=0.6\linewidth]{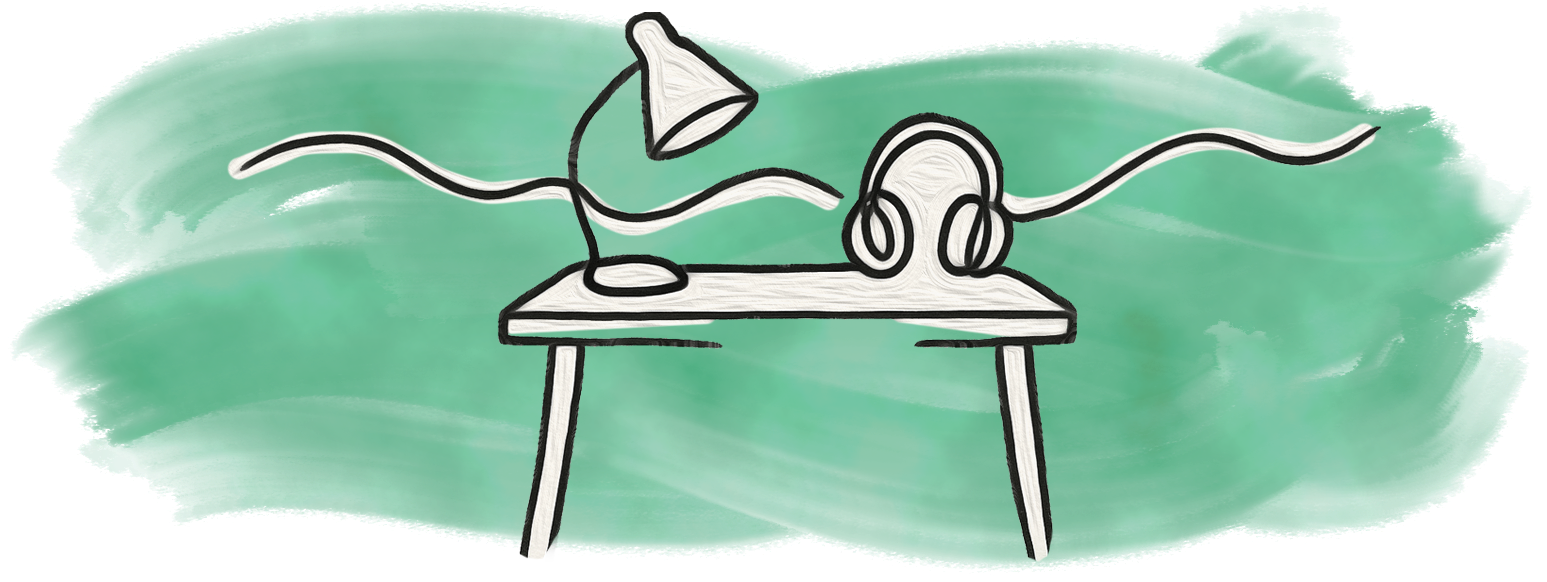}
    \caption{Optimizing the Environment}
    \label{fig:Optimizing the Environment}
\end{figure}

This category captures interventions through which participants sought to shape their physical or digital environment in order to create conditions conducive to flow. It comprises 13 distinct interventions, mentioned a total of 214 times (54\% of all mentions), making it the most frequently represented category in the dataset. A detailed overview over all mentioned interventions, their frequencies and power quotes can be found in Table~\ref{tab:environment}.

\begin{table*}[t]
\centering
\caption{Reported interventions for optimizing the environment, with counts and illustrative quotes.}
\label{tab:environment}

\begin{tabular*}{\textwidth}{p{3.8cm} r p{0.72\textwidth}}
\hline
\textbf{Intervention} & \textbf{\#} & \textbf{Participant quote} \\
\hline
reducing distractions & 60 & ``I turn off email and chat notifications during deep work periods to avoid distractions'' (P89) \\
background stimulus & 52 & ``Often I have calm music or calm background sound like a YouTube video on so a part of my brain is focused on that while the other part is fully focused on the work.'' (P112) \\
preparing the workspace & 30 & ``I make sure I have everything I might need within arms reach'' (P97) \\
quietness & 22 & ``Sit in a quiet area'' (P119) \\
wearing headphones & 17 & ``I use my headphones without music or sounds'' (P84) \\
being in a dedicated place & 10 & ``I often enter this state when I work remotely from home.'' (P96) \\
being alone & 8 & ``I am always alone when I want to be in flow.'' (P83) \\
comfort & 7 & ``I sit down comfortably'' (P125) \\
being among others & 2 & ``The fact that everyone else in the library is also working next to me has a big influence on me and motivates me.'' (P74) \\
performant IT system & 2 & ``A fast computer is absolutely necessary.'' (P130) \\
daylight & 2 & ``Enough daylight'' (P40) \\
fitting weather & 1 & ``Good weather'' (P48) \\
fresh air & 1 & ``I keep my terrace door open to get \dots\ a light breeze'' (P96) \\
\hline
\end{tabular*}

\end{table*}


The most frequently mentioned intervention, both in this theme and overall, centered on removing or minimizing distractions (n=60). These were often expressed in very literal terms (e.g., “I remove distractions.” - P105), but participants frequently specified email or chat notifications as key obstacles: “I turn off email and chat notifications during deep work periods to avoid interruptions.” (P89). Phones (n=29) and notifications from workplace IT systems (n=7) were the most commonly named sources of distraction , with participants using tools such as “Do Not Disturb” or “Airplane Mode” to silence them. Headphones (n=17), often with active noise cancellation (n=7), were another widely used device to shield off environmental interruptions. Because of the possibility that headphones serve additional benefits, beyond reduction of outside distractions, we kept them as a code distinct from (yet overlapping with)  "minimizing distractions".

Interestingly, the importance of minimizing distractions seemed to transition partly into the value of controlled background stimulation. Some participants framed music or other media as a way to counteract excessive unwanted noise: “I cannot focus when there is excessive noise or numerous distractions, I enjoy listening to music or a comfort show in the background as white noise.” (P95). Among those who deliberately used background stimulus (52 mentions), the majority (48) referred to music. For some, music primarily served as an acoustic shield (e.g. “With some music to cancel all the external sounds that could distract me.” (P152)), while others emphasized its role in regulating mood and energy (e.g. “listening to music that keeps me in a good mood.” (P144); “listening to music that has a calming effect.” (P22)). Participants also mentioned podcasts, ASMR content, TV shows, or nature sounds, for instance: “Keeping TV on with a low volume tends to calm me and gives a nice background.” (P96). Within music use, people did not mention actual tools they use, but often referred to curated playlists, with ambient, lo-fi, and classical genres most frequently named.

A third strand of interventions involved preparing the workspace in advance (n=30). This took two main forms: ensuring tidiness by removing unnecessary items (e.g., “Making sure my desk/work space is clean and not cluttered with things.” - P112), and ensuring availability by keeping all necessary materials close at hand (e.g., “I make sure I have everything I might need within arms reach.” - P97). Both aspects applied to physical as well as digital work environments.

\subsubsection{Planning and Organization}

\begin{figure}[H]
    \centering
    \includegraphics[width=0.6\linewidth]{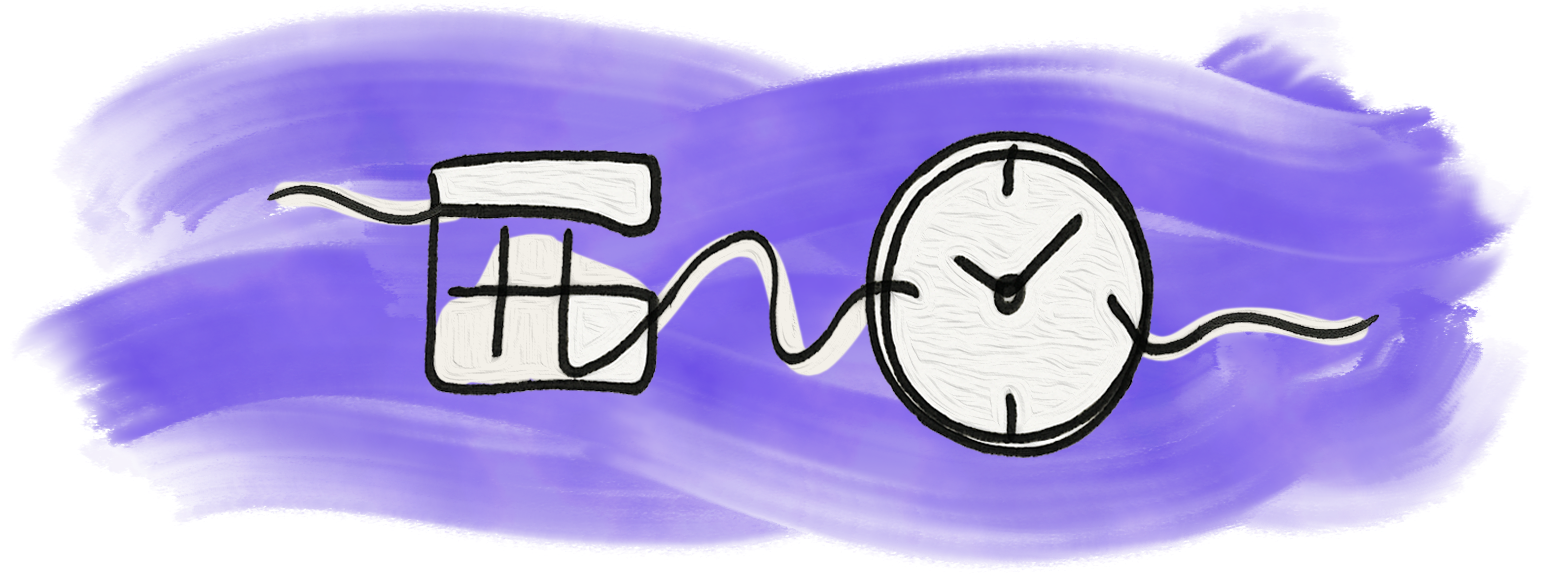}
    \caption{Planning and Organization}
    \label{fig:Planning and Organization}
\end{figure}

This theme encompasses strategies through which participants deliberately structured their work in time and routine to create conditions favorable for flow. It comprises 10 distinct codes, mentioned a total of 60 times (15\% of all mentions). This theme highlights how temporal structuring and cognitive externalization can play a crucial role in fostering flow. A detailed overview over all mentioned interventions, their frequencies and power quotes can be found in Table~\ref{tab:organization}.

\begin{table*}[t]
\centering
\caption{Reported interventions for planning and organizing work, with counts and illustrative quotes.}
\label{tab:organization}

\begin{tabular*}{\textwidth}{p{3.8cm} r p{0.72\textwidth}}
\hline
\textbf{Intervention} & \textbf{\#} & \textbf{Participant quote} \\
\hline
time boxing & 19 & ``Set a time frame for the work session'' (P44) \\
time pressure & 11 & ``Deadline, how close is it and how fast do I need to go to finish my task'' (P110) \\
specific time of day & 9 & ``Starting right after breakfast or lunch, because that's when I'm most focused'' (P36) \\
cognitive offloading & 8 & ``Thoughts put down in writing'' (P17) \\
regular breaks & 5 & ``I sometimes do quick pauses to energize and rest a bit but without losing focus'' (P96) \\
progress tracking & 3 & ``Track each task I complete on a dedicated file I created'' (P96) \\
patience / easing in & 2 & ``It sometimes just takes time to get into it'' (P41) \\
pre-work ritual & 1 & ``Create a pre-work ritual (much like a positive trigger)'' (P1) \\
no upcoming appointments & 1 & ``No upcoming appointments'' (P13) \\
routine & 1 & ``Routine'' (P18) \\
\hline
\end{tabular*}
\end{table*}

The most frequently mentioned strategy was time boxing (n=19), where participants scheduled fixed time slots for focused work. A typical description was: “Set a time frame for the working session.” (P44). Several participants reported using Pomodoro timers (n=3) to achieve this structure, which in turn probably also supported the related intervention of taking regular breaks (n=5).

Another commonly described approach was relying on time pressure (n=11). This was mostly named very succinctly as “time pressure.” (P2). Participants especially noted that deadlines provided a natural push toward flow, while others emphasized the usefulness of self-imposed, artificial deadlines as a way to induce the same effect.

Participants also emphasized the role of working at specific times of the day (n=9). One framed this more generally as the need to “identify productive hours” (P87), while others were more concrete: “start early in the morning.” (P5); “I start right after breakfast or lunch, then I am most concentrated.” (P36); or, conversely, “I often work at night.” (P83). These accounts underscore the highly diverse nature of reported personal temporal rhythms in supporting flow.

A further frequently mentioned intervention involved cognitive offloading (n=8). Participants described how freeing the mind from unrelated thoughts helped maintain immersion, often by writing them down: “thoughts put down in writing.” (P17). Relatedly, while mentioned less often, progress tracking (n=3) was described as helpful for keeping focus and motivation, usually through checklists or files dedicated to completed tasks: “track each task I complete on a dedicated file I created.” (P96).

\subsubsection{Shaping the Task}

\begin{figure}[H]
    \centering
    \includegraphics[width=0.6\linewidth]{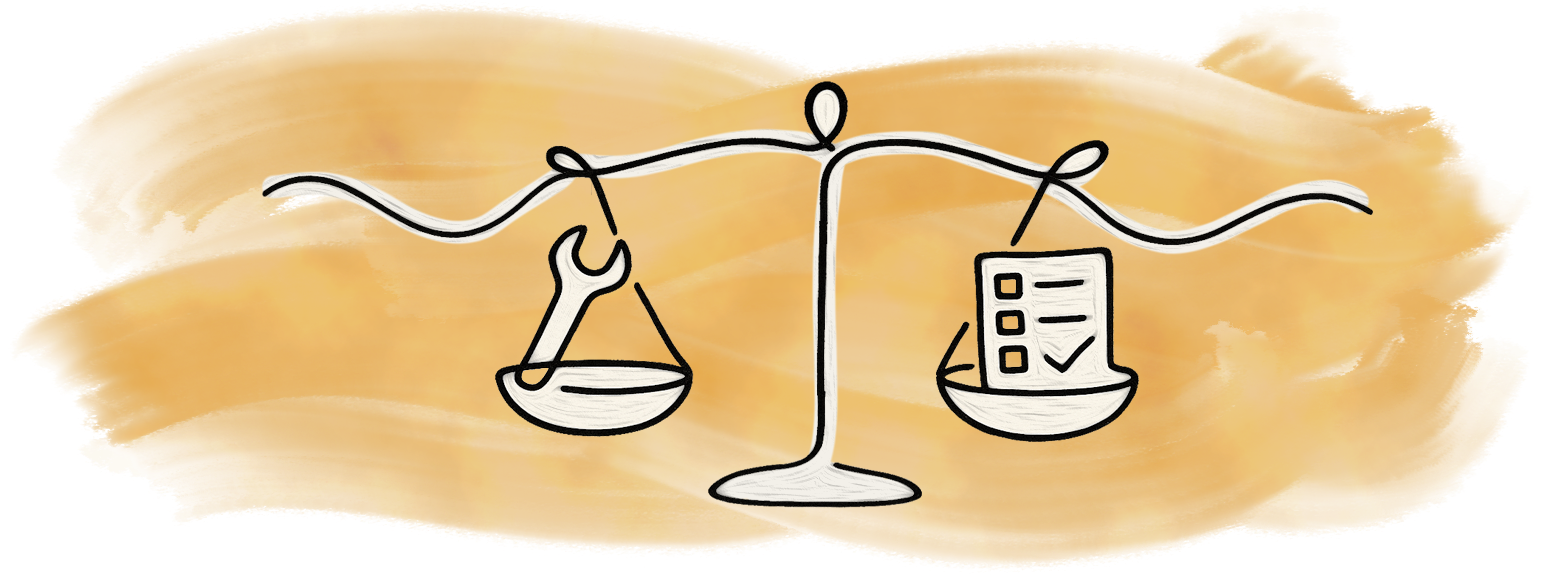}
    \caption{Shaping the Task}
    \label{fig:Shaping the Task}
\end{figure}

This theme captures strategies through which participants shaped the immediate work content itself, rather than the broader temporal or organizational context. It comprises five distinct codes, mentioned a total of 58 times (15\% of all mentions). While it is the theme with the fewest distinct codes, it still keeps pace with the others in terms of overall mentions, reflecting its centrality to participants’ lived experiences of flow. A detailed overview over all mentioned interventions, their frequencies and power quotes can be found in Table~\ref{tab:task}.

\begin{table*}[t]
\centering
\caption{Reported interventions for shaping the task, with counts and illustrative quotes.}
\label{tab:task}

\begin{tabular*}{\textwidth}{p{3.8cm} r p{0.72\textwidth}}
\hline
\textbf{Intervention} & \textbf{\#} & \textbf{Participant quote} \\
\hline
clear goals & 28 & ``Set clear tasks and write them down for myself on a piece of paper'' (P27) \\
meaningful task & 13 & ``If I find the topic exciting, I quickly and easily get into a flow. If I'm not so interested in the topic, I find it particularly difficult.'' (P9) \\
skill--demand balance & 10 & ``The most important prerequisite is an appropriate level of difficulty—not too hard, not too easy.'' (P62) \\
single-tasking & 4 & ``I avoid multitasking by dedicating full attention to one dataset or analysis at a time'' (P89) \\
task prioritization & 3 & ``Prioritize the harder things first and get them done while my mind is still fresh'' (P95) \\
\hline
\end{tabular*}
\end{table*}


The most frequently mentioned intervention was setting clear goals (n=28). Participants highlighted how clarity about objectives directly enabled immersion: “Setting goals – I set for myself clear goals which are realistic and try by all means to achieve them.” (P142). Others emphasized the importance of receiving unambiguous work assignments: “The most important thing is a clear work assignment! If you know what you have to do, then you can just do it... If you first have to figure it out and aren't sure whether you're doing the right thing, you won't get into the FLOW at all.” (P67). Several participants described using checklists (n=8) as a simple tool to create such goals. While they are similar tools as to-do-lists, which were named for tracking progress, they serve a second purpose for this interventions: to actively create and then follow these goals to work towards.

Another frequently mentioned factor was working on meaningful or interesting tasks (n=13). Many participants framed this as a precondition for flow, for instance: “I know that I need to be working on something meaningful and fresh.” (P170). At the same time, some participants described how they actively created meaning. One example was: “Write down for myself what I expect to gain from the learning session. I throw these notes away after the session, but thinking about what I will achieve by completing a task gives me a lot of motivation.” (P27).

Participants also referred to maintaining a balance between task demands and skills, consistent with flow theory. This was often described as a prerequisite: “The most important prerequisite is an appropriate level of difficulty—not too hard, not too easy.” (P62). Yet some participants shared strategies to actively manage this balance. For example, one described self-assessing skills at the outset of a session: “How good am I really? To answer that question can get me in the flow state.” (P110). Others emphasized longer-term approaches akin to job crafting, such as focusing on activities that challenge skills (P109). Participants also reported ways to handle imbalance, either by seeking help when tasks felt overwhelming (P157) or delegating overly simple tasks (P120).

Finally, single-tasking (n=4) and task prioritization (n=3) appeared as two additional less frequently named interventions. Although these strategies also relate to planning and organizing, we included them here because they describe how participants directly shape their engagement with the task at hand.

\subsubsection{Ensuring Mental and Physical Readiness}

\begin{figure}[H]
    \centering
    \includegraphics[width=0.6\linewidth]{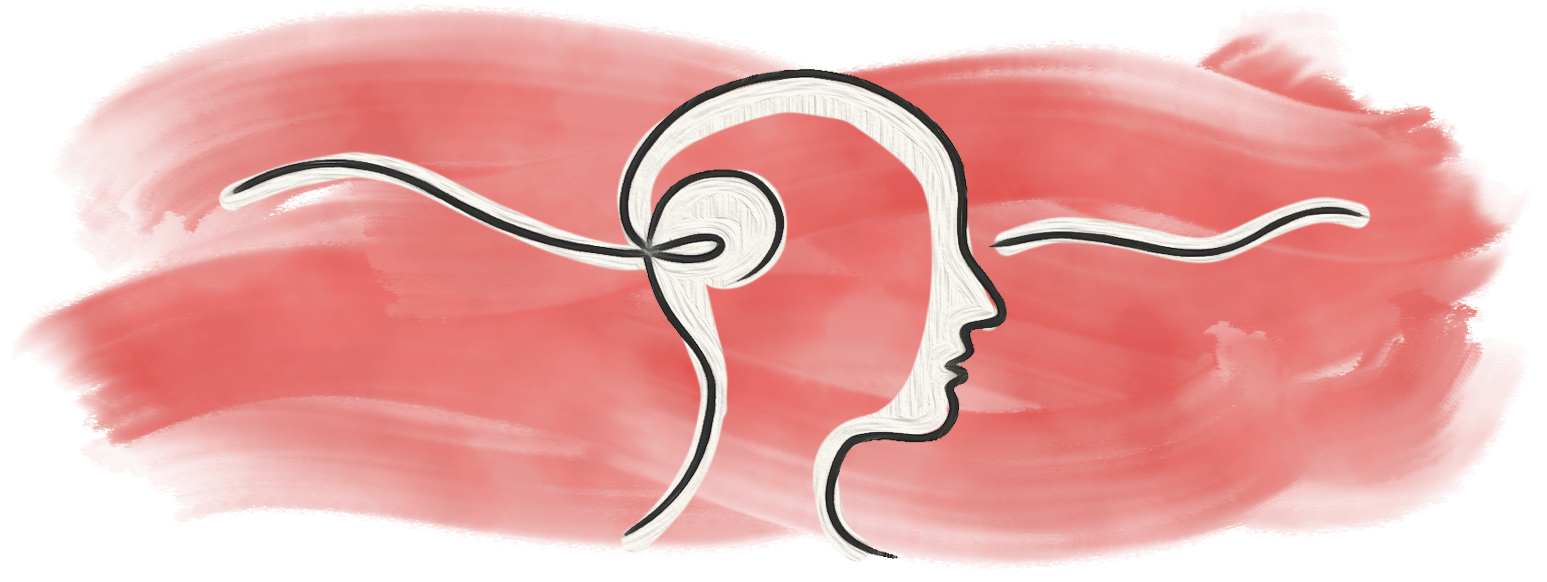}
    \caption{Ensuring Mental and Physical Readiness}
    \label{fig:Ensuring Mental and Physical Readiness}
\end{figure}

This theme encompasses strategies through which participants supported their flow by attending to their physical and mental state. It comprises ten distinct codes, mentioned a total of 63 times (16\%). While diverse in nature, these interventions share the idea that flow depends not only on external or task-related conditions but also on being personally ready to engage. A detailed overview over all mentioned interventions, their frequencies and power quotes can be found in Table~\ref{tab:readiness}.

\begin{table*}[t]
\centering
\caption{Reported interventions for ensuring mental and physical readiness, with counts and illustrative quotes.}
\label{tab:readiness}

\begin{tabular*}{\textwidth}{p{3.8cm} r p{0.72\textwidth}}
\hline
\textbf{Intervention} & \textbf{\#} & \textbf{Participant quote} \\
\hline
well nourished & 13 & ``A healthy and well-balanced diet'' (P80) \\
physical activity & 11 & ``I exercise before work'' (P108) \\
stimulants & 10 & ``Get myself a cup of coffee before working'' (P149) \\
mindfulness and meditation & 10 & ``Meditation, in case you lose your flow and need to clear your head to continue'' (P5) \\
sufficient sleep & 10 & ``Having slept well'' (P48) \\
general well-being & 4 & ``No physical problems (back pain, headaches, \dots)'' (P3) \\
self-motivation & 2 & ``Self-motivation'' (P8) \\
right mindset & 1 & ``Right mindset'' (P46) \\
self-conversation & 1 & ``Self conversation'' (P153) \\
discipline & 1 & ``Discipline'' (P61) \\
\hline
\end{tabular*}
\end{table*}


The most frequently named intervention was being well nourished (n=13), which included both food and hydration. Participants emphasized that an appropriate nutritional state before working, as well as access to drinks or snacks during work, was important for staying immersed: “not hungry and not too full.” (P37); “drink a lot of water while working.” (P118). Several participants explicitly noted the value of healthy food choices to sustain energy.

Physical activity (n=11) was another prominent intervention. Similar to nutrition, exercise was mentioned both as a preparatory activity and as an ongoing routine: “I exercise before work.” (P108); “very regular exercise.” (P80). Relatedly, participants also highlighted the importance of being well rested (n=10) and maintaining general health (n=4) as prerequisites for entering and sustaining flow.

The use of stimulants (n=10), particularly caffeine, was another frequently reported practice. Participants described coffee and similar drinks as reliable tools to increase alertness and support concentration. In contrast to such physiological boosts, others turned to mental practices: mindfulness and meditation (n=10) were often mentioned as ways to foster readiness. Some participants used them at the start of a work session (“I start with meditation, exercises, mindfulness.” (P98)), while others described them as interventions during breaks when focus waned: “Meditation, in case you lose your flow and need to clear your head to continue.” (P5).

\subsection{\revision{Interim Discussion}}

Study 1 yielded a diverse repertoire of 38 lived interventions across four categories, reported by a sample of 80 university students and 80 digital workers from a broad range of countries and age groups. Taken together, these accounts provide a broad picture of how digital workers deliberately foster flow. Due to the wide variety of interventions and experiences captured, and the validation of our initial structure with the second sample, we consider the repertoire to be sufficiently rich in breadth. 

The frequency of mentions we report offers an initial indication of which interventions are more salient, yet this metric must be interpreted carefully. Because participants were prompted only once to freely recall strategies, responses are shaped by availability bias \cite{tversky1973availability}: easily retrievable strategies are more likely to be reported first, while less salient but potentially effective interventions may remain underrepresented. Thus, raw mention counts reflect prominence in memory more than actual effectiveness. Similarly, the absence of a strategy in a participant’s response does not imply that it is unhelpful—it may simply not have been tried, or not have come to mind in that moment. Another limitation is that Study~1 cannot capture polarization in how interventions are experienced. A strategy might be highly beneficial for some individuals but distracting or even counterproductive for others. Our free-response design surfaces breadth, but it does not provide systematic information about such conflicts in endorsement. 

Taken together, these considerations highlight that while Study~1 reveals a rich collection of strategies digital workers employ, more research is needed to fully assess our RQ1. Addressing it requires a systematic follow-up that evaluates the repertoire across a new participant sample, thereby reducing availability bias and probing which interventions are widely endorsed, contested, or disendorsed. We turn to this next in Study~2.
\section{Study 2: Validating and Differentiating the Intervention Landscape}

Building on Study~1, which surfaced a repertoire of lived interventions from open-ended responses, Study~2 presented this structured list to a new sample in order to systematically capture endorsement and disendorsement. The goal was to address the availability bias inherent in free recall and to examine not only which interventions are broadly endorsed, but also where opinions diverge and potential conflicts emerge. In this way, Study~2 provides a more differentiated picture of how digital workers pursue flow.

\subsection{Method}

\begin{figure*}[t]
    \centering
    \includegraphics[width=1\linewidth]{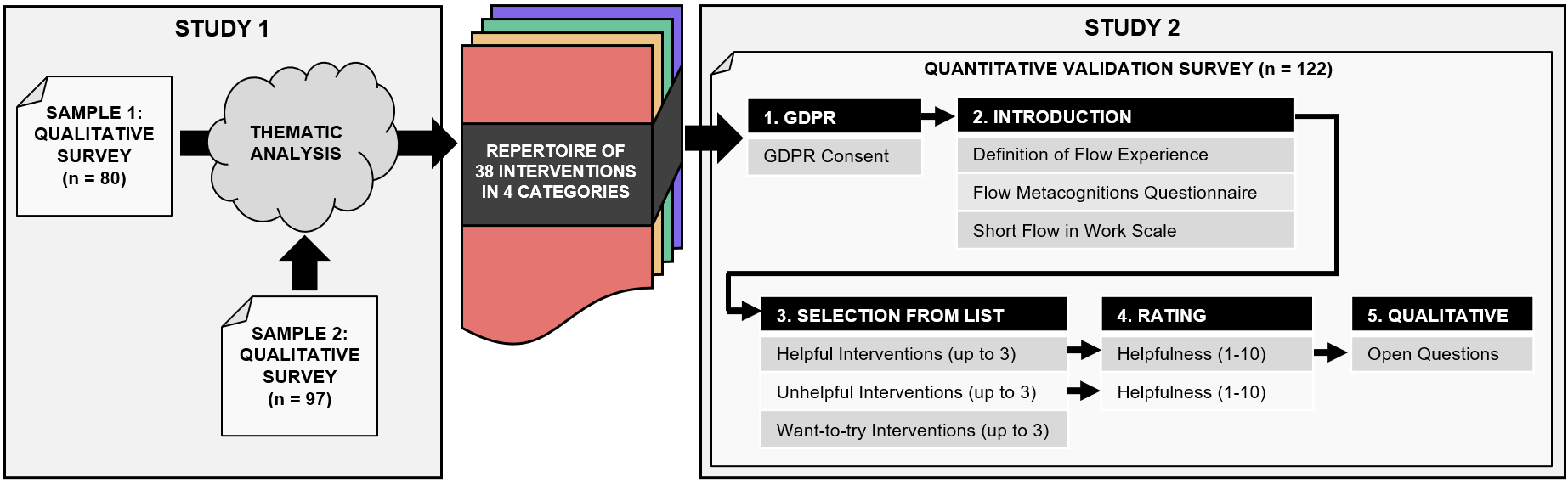}
    \caption{Overview of the Procedure, showing both studies. Study~1 consisted of two samples of the same qualtiative survey, which were sequentially analyzed through thematic analysis. The resulting repertoire of 38 Interventions in 4 Categories was the basis for Study~2. In Study~2, participants first completed flow questionnaires, then evaluated the repertoire of interventions on helpfulness, unhelpfulness, and curiosity, followed by ratings and open-ended elaborations.}
    \label{fig:procedure}
\end{figure*}

\subsubsection{\revision{Study Design}}

Study 2 was conducted as an online survey experiment implemented in oTree \cite{chenOTreeOpensourcePlatform2016a}.
After providing informed consent in compliance with GDPR regulations, participants completed the study in four main parts:

Participants first completed two flow-related questionnaires. As in Study 1, they were introduced to the flow construct and filled in the Flow Metacognition Questionnaire \cite{wilsonFlowMetacognitionsQuestionnaire2016a} together with the Short Flow in Work Scale \cite{monetaValidationShortFlow2017a}, which together assessed their general tendency to experience flow at work, and their attitudes towards the flow experience (for details see Table~\ref{tab:questionnaires}).

Next, they were presented with the list of interventions derived from Study 1. They were asked to indicate up to three strategies that they considered helpful for experiencing flow, up to three they had tried but found unhelpful, and up to three they had not yet tried but would like to. We capped the number of selections at three for two reasons: first, this reflects the rounded average number of strategies reported in Study 1 (2.47), and second, each selected strategy triggered additional follow-up questions, making completion time increase disproportionately with the number of selections. To ensure participants could give thoughtful answers without the survey becoming overly long, we restricted the maximum number of selections. To reduce order effects, the list was randomized individually for each participant.

For interventions marked as helpful and tried but unhelpful, participants provided a helpfulness rating on a 10-point scale (1 = not at all helpful, 10 = very helpful). Finally, participants were invited to elaborate in two optional open-ended questions for each helpful strategy, describing how it supported their flow at work and how a technological system (e.g., an app or device) might provide support for this intervention. An overview of the procedure of Study~2, and how it is set in the whole research design is displayed in Figure~\ref{fig:procedure}.

\subsubsection{Participants}
To accommodate participants from diverse cultural backgrounds and time zones, the study was available for a 24-hour period and distributed across multiple sub-studies \revision{on Prolific}. Sub-studies were targeted at six world regions (North America, South America, Europe, Africa, South-East Asia \& Oceania, Rest of Asia) to ensure broad geographical representation. Within each sub-study, a 50/50 sex quota was applied to achieve balanced representation. Eligibility criteria required participants to be fluent in English, employed either full-time or part-time, and spend more than 75\% of their daily work time on computers. Individuals who had taken part in Study 1 were excluded from participation. \revision{As in Study 1, we tried to ensure to only record data from individuals who experience flow at work at least sometimes (for details, see Section 3.1.1.). Participants were compensated for their participation with £2.50.}
126 participants (64 female, 62 male) from 23 countries\footnote{\revision{The countries of our respondents were Australia, Austria, Brazil, Canada, Chile, Czech Republic, France, Germany, Greece, India, Indonesia, Italy, Kenya, Mexico, Morocco, New Zealand, Poland, Portugal, South Africa, Spain, UK, USA and Vietnam.}} \revision{responded to our survey}, with a mean age of 35.68 years (SD = 10.76, range = 20–65). A majority of participants (n = 105) reported full-time employment, while 22 were part-time employed. 

\revision{
\subsubsection{Data Analysis}
All quantitative analyses for Study~2 were conducted in R (2024.12.0 Build 467). After data cleaning and quality checks (see Section~4.2.1), we computed aggregate metrics (see Table \ref{tab:metrics}) to characterize participants’ evaluations of each intervention.
We then used these metrics to compare interventions along key dimensions of endorsement and agreement. 
To identify broader data patterns we applied k-means clustering on the dimensions helpfulness and polarization, selecting the number of clusters via elbow and silhouette analyses.
Finally, to examine whether individual differences in flow metacognition shaped intervention preferences, we divided participants into high and low FMQ\textsubscript{2} groups using a median split. We descriptively compared endorsement frequencies across these groups and visualized the resulting patterns in Figure~\ref{fig:fmq_endorsements}. We do this to uncover whether people with a higher ability to regulate their own flow use other interventions than people with a lower ability (or vice versa).
In addition to the quantitative analyses, we analyzed the open-ended responses in which participants described how technological support for their preferred interventions could be designed. We conducted a brief qualitative content analysis following a descriptive, low-inference approach. Our goal was not to produce a full thematic model but to summarize participants’ envisioned forms of tool support. In presenting these ideas, we report both recurring patterns across participants and particularly novel individual suggestions when they offered distinctive or illustrative perspectives. This analysis provides additional context for understanding how workers imagine technology could augment their lived interventions.}

\subsection{Results}
Study~2 revealed clear differences in how participants evaluated the repertoire of lived interventions. Some strategies were broadly endorsed, others showed strong disagreement across workers, and a few emerged as niche practices valued only by certain individuals. The data collected from this study are available in the supplementary material.


\begin{table*}[t]
\centering
\caption{Aggregate metrics derived from Study~2 used in the quantitative analysis of interventions.}
\label{tab:metrics}

\begin{tabular*}{\textwidth}{p{0.1\textwidth} p{0.4\textwidth} p{0.5\textwidth}}
\hline
\textbf{Metric} & \textbf{Definition} & \textbf{Interpretation} \\
\hline
RM\textsubscript{total} &
Mean of all ratings of an intervention by participants who selected it as helpful or not helpful &
Overall perceived helpfulness of an intervention across all \newline respondents \\

RM\textsubscript{helps} &
Mean of ratings of an intervention only by participants who selected it as helpful &
Perceived helpfulness of an intervention among endorsers \\

Polarization &
$1 - \left|\dfrac{n_{\text{helps}} - n_{\text{helps-not}}}{n_{\text{helps}} + n_{\text{helps-not}}}\right|$ &
Extent to which an intervention was simultaneously \newline experienced as helpful and not helpful \\

FMQ\textsubscript{1} &
Mean of FMQ items 1, 3, 5, 7, 9, and 11 &
Belief that the flow state fosters achievement \\

FMQ\textsubscript{2} &
Mean of FMQ items 2, 4, 6, 8, 10, and 12 &
Confidence in one’s ability to regulate flow \\

FMQ &
Mean of all FMQ items (1--12) &
General level of flow metacognition \\
\hline
\end{tabular*}
\end{table*}

\subsubsection{Sample Description \& Quality Checks}
We applied additional quality checks to identify participants whose responses suggested inattentive or patterned answering, while still making it past the easy attention checks.
For that, we (1) computed the average list position of selected items, (2) checked for the mean ratings of unhelpful interventions, and (3) computed the total word count of responses to the open questions. If the cumulative probability of (1) happening was <1\%\footnote{The probability was estimated via Monte Carlo simulation (100{,}000 samples), drawing the number of interventions that the participant selected from the full set of 38 interventions. We then computed the proportion of simulated draws with a mean list position lower than the participant’s observed mean. Very low probabilities (<.01) indicate that such a mean would be highly unlikely if selections were made deliberately, suggesting that the participant may have simply chosen the first few items.}, (2) was > 5 and (3) was 0, we excluded the participant from further analysis. This led to the exclusion of 5 participants. The metrics we will use in further analysis are explained in detail in Table~\ref{tab:metrics}.

\begin{figure*}
    \centering
    \includegraphics[width=0.8\linewidth]{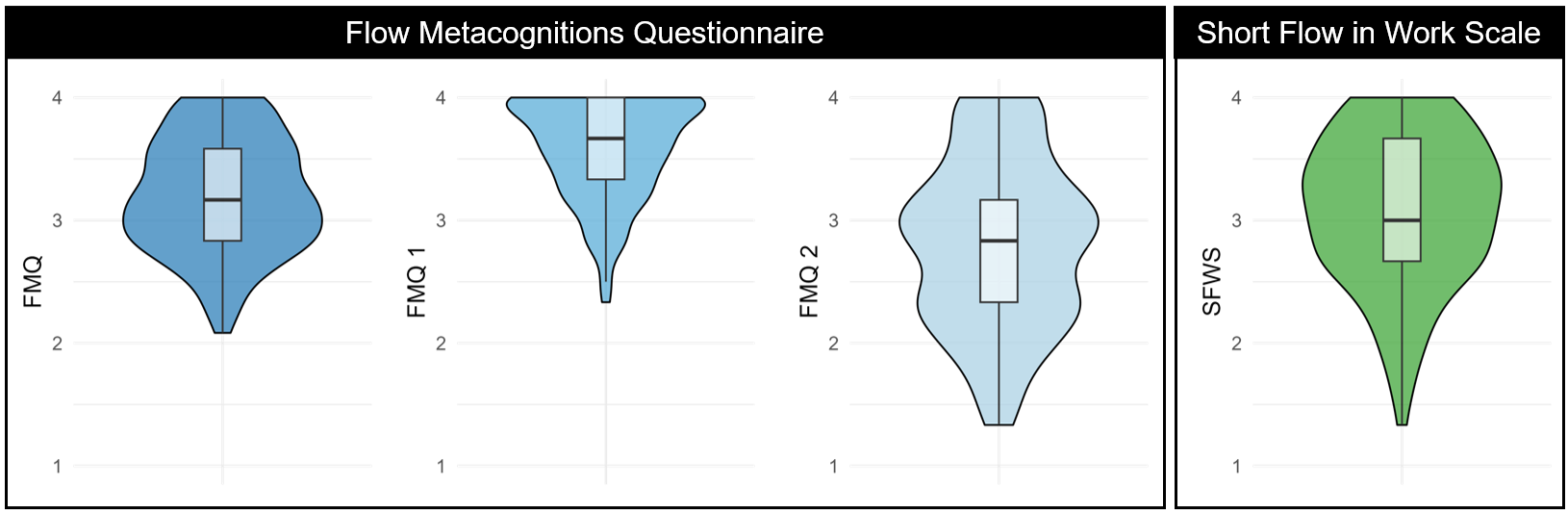}
    \caption{Distribution of Flow Metacognition Questionnaire (FMQ) and Short Flow in Work Scale (SFWS) scores. FMQ\textsubscript{1} (beliefs about flow benefits) was higher on average than FMQ\textsubscript{2} (confidence in regulating flow).}
    \label{fig:fmq_sfws}
\end{figure*}

Flow Metacognition scores were quite high on average 
($\mu_{\text{FMQ}} = 3.20$, $SD_{\text{FMQ}} = 0.46$), 
mainly driven by the first factor 
($\mu_{\text{FMQ}_{1}} = 3.59$, $SD_{\text{FMQ}_{1}} = 0.41$). 
The factor reflecting the ability to regulate flow experiences 
was lower on average 
($\mu_{\text{FMQ}_{2}} = 2.80$, $SD_{\text{FMQ}_{2}} = 0.68$). The SFWS score had a mean of 3.10 (SD = 0.61).
For detailed information see Figure~\ref{fig:fmq_sfws}.

\subsubsection{Selection Distributions}
\begin{figure*}
    \centering
    \includegraphics[width=1\linewidth]{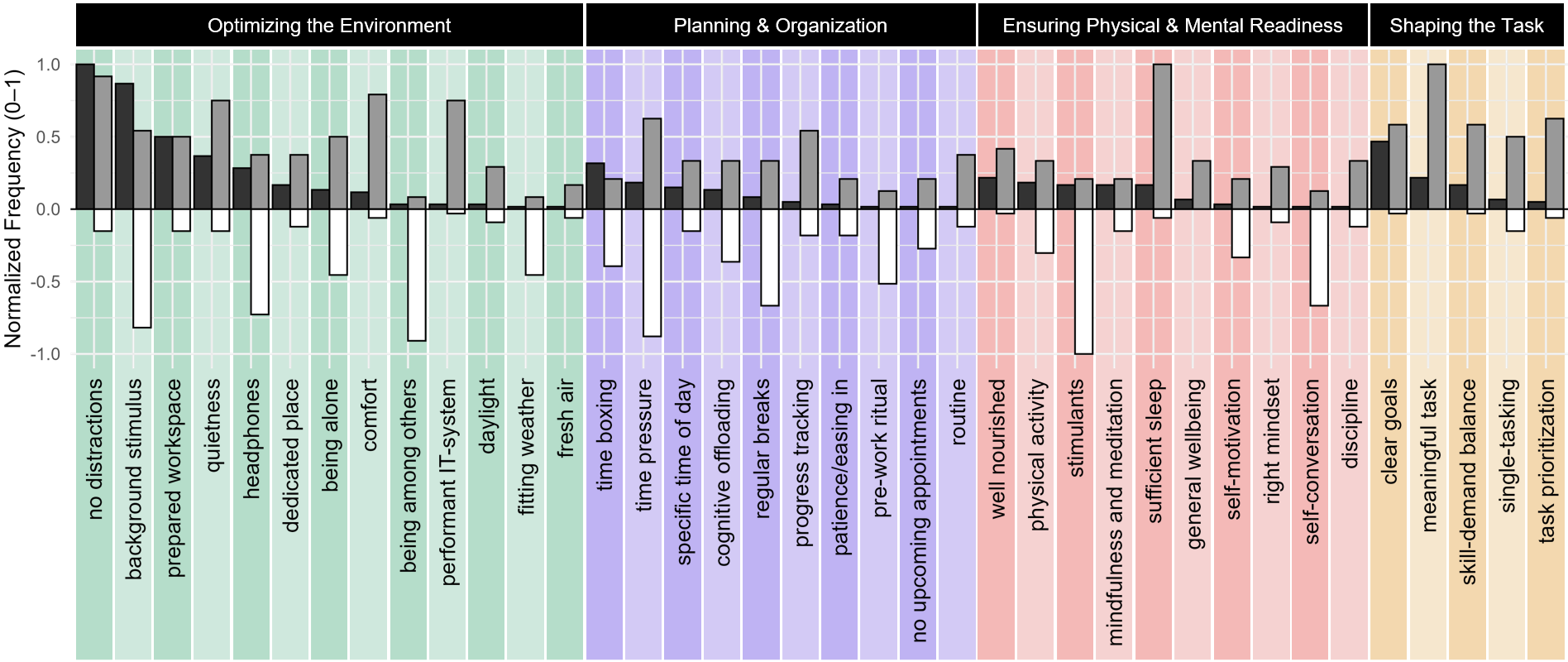}
    \caption{Comparison of interventions across studies. Black bars show raw frequencies in Study~1. Grey bars show counts of helpful selections and white bars show counts of unhelpful selections in Study~2, highlighting consensus or polarization.}
    \label{fig:interventions_bar}
\end{figure*}

Across the 122 participants, we observed a large number of selections: on average, they selected 2.96 interventions as \emph{helpful}, 2.76 interventions as \emph{unhelpful}, and 2.84 interventions as \emph{to-try}. We first compared these frequencies with the raw frequencies of mentions obtained in Study~1. To visualize emerging patterns, we prepared a grouped bar plot (see Figure~\ref{fig:interventions_bar}) that displays, for each strategy, its raw frequency of mentions in Study~1, the number of \emph{helpful} selections in Study~2, and the number of \emph{unhelpful} selections in Study~2. 
At a high level, this comparison reveals interesting differences in distribution. Several interventions that were highly salient in Study~1 (e.g., background stimulus, distraction blocking) were also among the most frequently chosen as \emph{helpful} in Study~2. However, the structured format also highlighted polarization effects: certain interventions (e.g. background stimulus, time pressure) emerged as both supportive for some participants and not helpful for others.

\begin{figure*}
    \centering
    \includegraphics[width=1\linewidth]{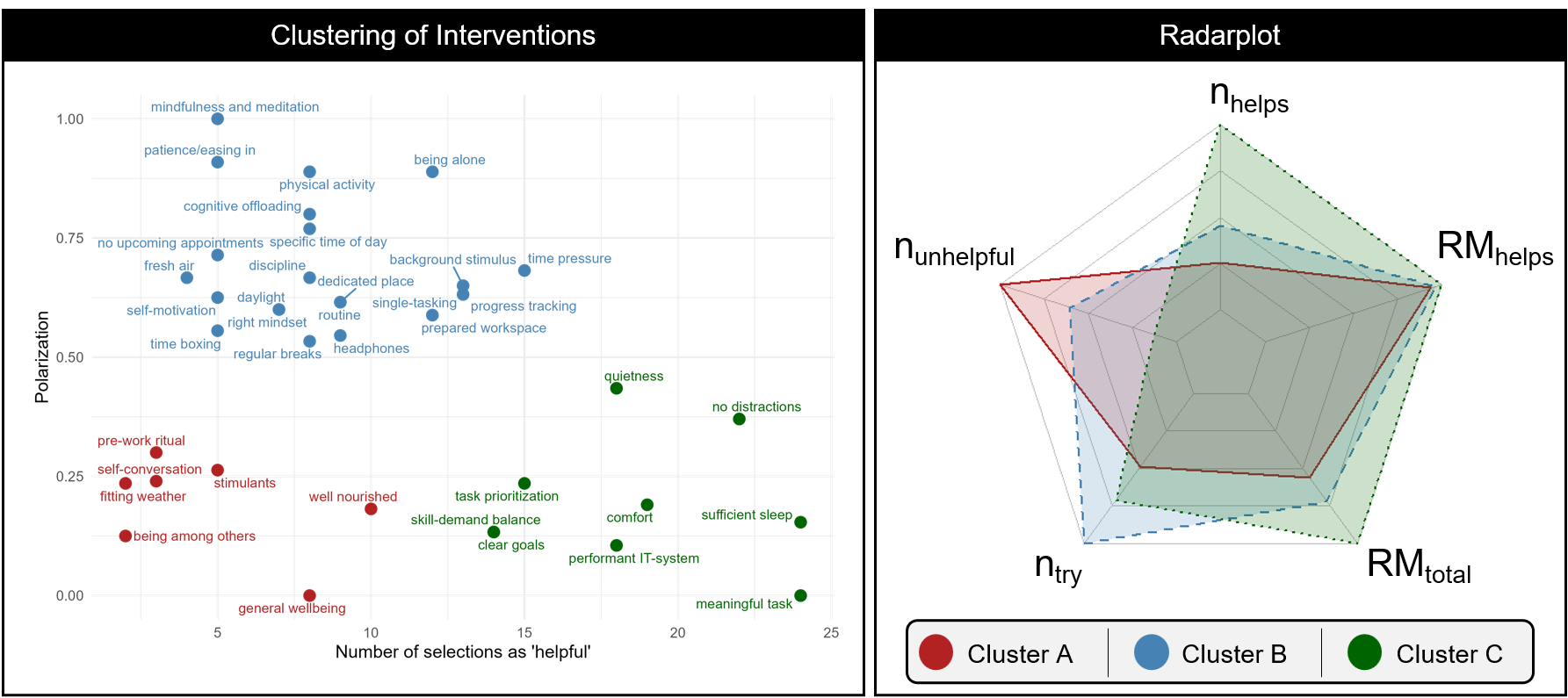}
    \caption{Clustering of interventions in a two-dimensional space of helpfulness and polarization. The three-cluster solution differentiates niche (A), controversial (B), and widely endorsed (C) strategies.}
    \label{fig:clusters}
\end{figure*}

To compare interventions in terms of their effectiveness for fostering flow, we mapped them in a two-dimensional space defined by (a) the number of participants selecting them as helpful and (b) a polarization score capturing user agreement (0 = high agreement, 1 = high disagreement). We applied k-means clustering and determined the number of clusters by comparing solutions for k = 2–10 using the elbow method and silhouette analysis. The within-cluster sum of squares dropped sharply until k = 3 (18.55), after which gains leveled off. Silhouette widths likewise peaked at k = 3 (0.54), outperforming k = 2 (0.49) and all k $geq$ 4 (<0.44). Based on this convergence of evidence, we adopted a three-cluster solution (Figure~\ref{fig:clusters}). The resulting clusters represent distinct types of interventions:

\begin{enumerate}
    \item[A] \textbf{Niche interventions:} Interventions that were rarely selected and not broadly perceived as helpful, yet still used successfully by a few participants. Among the clusters, this one is characterized by the highest number of selections as unhelpful and the lowest selections as helpful or “want to try.” Nevertheless, the high RM\textsubscript{helps} scores show that for those who do find them useful, they are just as effective as interventions in the other clusters. In contrast, RM\textsubscript{total} scores are lowered by the negative ratings of participants who tried these interventions but did not find them beneficial. This suggests that such interventions are not universally effective and should be approached with caution. Examples in this cluster include using stimulants (e.g., caffeine) or working in the presence of others.
    
    \item[B] \textbf{Controversial interventions:} Interventions that were helpful for some participants but ineffective for others. These clearly require personalization, and notably, they also attracted the highest number of “want to try” selections. With 21 interventions, this is the largest cluster. In terms of helpful and unhelpful counts as well as rating metrics, it sits between the other two clusters, but stands out through the strong curiosity and experimentation interest it generated. Taken together, these interventions show considerable potential when matched to the right individual. Examples from this cluster include creating time pressure, introducing background stimuli (e.g., music), or engaging in cognitive offloading. 
    
    \item[C] \textbf{Widely endorsed interventions:} Interventions that were commonly selected and associated with strong consensus on their helpfulness. This cluster is characterized by the highest number of selections as helpful and by leading scores on both rating metrics. Conversely, it also had the lowest number of unhelpful selections. Many task-focused interventions fall into this cluster, such as setting clear goals or working on a meaningful task.
\end{enumerate}

\subsubsection{Differences in Flow Metacognition}
\begin{figure*}[t]
    \centering
    \includegraphics[width=1\linewidth]{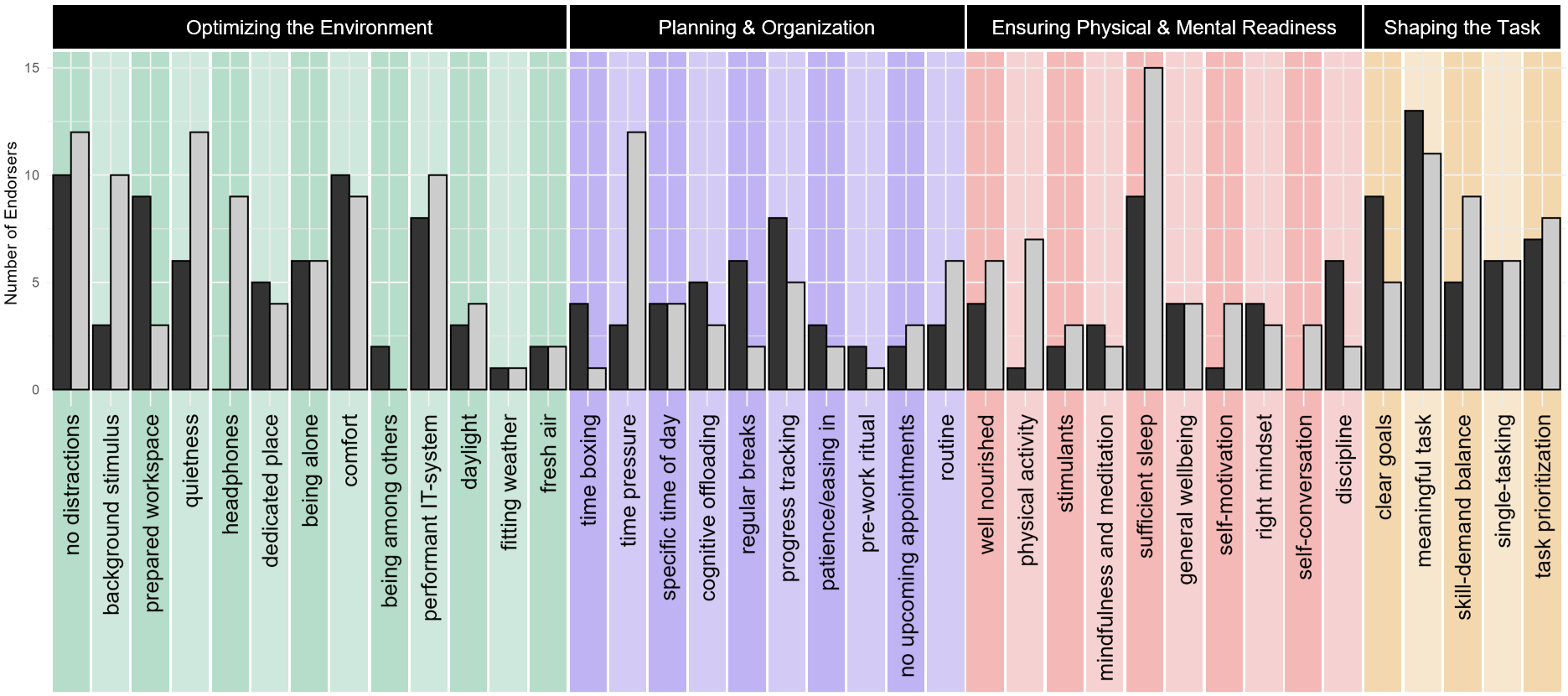}
    \caption{Differences in endorsement of interventions between participants with low vs.\ high FMQ\textsubscript{2} scores (confidence in regulating flow). Black bars indicate high FMQ\textsubscript{2}, and grey bars indicate low FMQ\textsubscript{2}.}
    \label{fig:fmq_endorsements}
\end{figure*}

To examine whether participants’ flow metacognition influenced which interventions they endorsed, we split participants into high versus low groups based on a median split of the FMQ\textsubscript{2} construct (confidence in ability to regulate flow). For each intervention, we then counted endorsements separately for high- and low-FMQ\textsubscript{2} participants \revision{and visualize this distribution in} Figure~\ref{fig:fmq_endorsements}.

\revision{We examined the most convergent and divergent response patterns between the low- and high-FMQ\textsubscript{2} groups.}
Endorsement patterns were largely similar across confidence levels, with comparable intervention selection rates.
However, \revision{several interventions showed larger imbalances of five or more selections, including \emph{headphones}, \emph{time pressure}, \emph{background stimulus}, \emph{sufficient sleep}, \emph{physical activity}, \emph{quietness}, and \emph{prepared workspace}. All of these were endorsed more often by participants with lower flow-metacognition confidence, with the exception of \emph{prepared workspace}, which was the only intervention in this subset more frequently chosen by high-FMQ\textsubscript{2} participants.}
\revision{These patterns are descriptive and must be interpreted cautiously due to the modest sample size, yet they suggest that workers with different levels of flow-regulation confidence may not benefit from the same interventions.}

\subsubsection{Participants’ Design Ideas}

\revision{Finally, participants could describe ideas how technological support for their preferred interventions could be designed. 
A first set of ideas focused on \emph{background calibration}, in which systems would adjust environmental inputs dynamically rather than leaving users to intervene manually. For example, (P2) imagined “a tool that could recognize when I'm in flow and then slowly fade out the music because I don't need it anymore would be cool”, while P(9) envisioned “a VR headset with noise-cancelling functions that…provide a more peaceful and solitary virtual space” in busy offices. 
Others ideated systems that automatically configure workspace conditions, for instance by “block[ing] distractions, adjust[ing] lighting and sound for focus, and track[ing] flow sessions” (P107). 
Some participants emphasized shaping work through \emph{meaning-led task support}. (P17) wished for automation of routine tasks so that cognitive capacity could be directed toward “the more interesting tasks”. 
Participants also described systems that help them \emph{find natural seams} in the workday. For example, (P98) asked for break cues aligned with well-being (“Let me know when to take a break so I can enjoy some daylight”), while (P2) imagined support for progress visibility that appears only when helpful: “a tool that could give me the to-do list exactly when I need it…or automatically track my progress” without interrupting flow. 
Several participants' answers highlighted the importance of \emph{reducing friction through automation}. (P72) described wanting a tool that “automatically block[s] off my calendar…[and] put[s] an automatic redirection on my incoming phonecalls” to maintain protected focus time.
Another group of ideas revolved around \emph{calibrated urgency}. Participants described systems that encourage gently bounded deadlines, such as one that “organises all of my tasks by date and encourages me to set deadlines…giving me the right frame to be productive” (P80), or tools that help create “good pressure” without inducing stress (P1). 
Participants also frequently imagined forms of \emph{sensing-guided} personalization that would allow tools to adapt based on momentary states or contextual cues. Across many ideas, sensing appeared as a meta-requirement for supporting flow, such as recognizing deep-focus states to fade out audio (P2), automatically configuring workspace conditions (P107), or even “measur[ing] my caffeine intake and alert[ing] me to when I need to consume more” (P15). 
Some participants envisioned systems, that deeply \emph{know the user and adapt to them}. (P116) wished for a system that understands their expertise so it can recommend suitable tasks (“If I could input my skills…[so it] could say what tasks I am confident working on”), while (P84) wanted help prioritizing tasks “based on importance and how good I am at completing them”. 
Finally, some participants wished for a digital mechanism to protect uninterrupted focus time in a \emph{socially visible} way. (P80) described time-boxed work as a “do not disturb period where I can just give everything to one task”, while (P14) wished for “an app that would stop my boss constantly interrupting me when I’m in my flow state”. 
Across these ideas, participants consistently envisioned flow-support systems that are highly personalized, reduce friction and routine burdens, and adapt automatically to the user’s context, skills, and moment-to-moment state. Taken together, these suggestions illustrate a preference for systems that learn about the user and intervene proactively to protect or restore conditions conducive to flow.}

\subsection{\revision{Interim Discussion}}
Study 2 allowed us to move beyond the availability bias inherent in Study 1 and systematically assess which interventions generalize, which polarize, and which remain niche. The resulting clusters have potential to sharpen our understanding of flow support: interventions in Cluster C emerge as broadly endorsed “safe bets” that can be considered baseline supports, while those in Cluster B highlight strategies that are highly promising for some but ineffective for others, and Cluster A interventions represent more niche practices. Taken together, this mapping frames the repertoire not as a universal toolbox but as a landscape with distinct zones of effectiveness. We therefore argue that flow support in HCI is best understood as an adaptation problem, rather than a fixed set of best practices. At the same time, these findings still need to be interpreted while keeping in mind that endorsement ratings capture perceived helpfulness, not objective effectiveness. Some strategies may be over- or under-valued due to expectations or prior experiences. Second, although we capped the number of selections to reduce fatigue, the sheer amount of interventions in the list may have limited the repertoire participants considered. 

Another critical implication is that higher polarization scores should not be dismissed as signals of poor interventions. Rather, they point to strategies that demand personalization. Indeed, the very potential of digital interventions may lie in these controversial Cluster B strategies, since they invite adaptation in ways that static, consensus strategies do not. System designers should thus approach Cluster C interventions (e.g., clear goals, meaningful tasks) as defaults that can be embedded broadly, while treating Cluster B interventions as opportunities for adaptive tuning and further experimentation.

Differences in FMQ\textsubscript{2} appear \revision{to matter only for certain interventions. The baseline ability to regulate flow may influence which strategies individuals perceive as helpful, suggesting that flow-support systems should consider these individual differences when recommending or tailoring interventions.} 

Finally, the large number of “want-to-try” selections underscores that cultivating flow often involves curiosity and self-experimentation. Workers actively explore strategies rather than relying solely on tried-and-true methods. This points to a design space for discovery tools that support such experimentation—for example, lightweight probes, recommenders, or reflection prompts that make it safe and visible for users to test new strategies and learn what works for them.
\section{\revision{Discussion}}

Our two studies surface a repertoire of lived flow interventions and show how they sort into zones of broad endorsement (Cluster~C), controversy (Cluster~B), and niche use (Cluster~A). Going further, we locate what HCI has already done within our four categories, tying these systems to flow theory. We then draw out what is missing based on our repertoire, and evaluate which gaps look most promising. Finally, we pair these opportunities with participants’ envisioned tool support from Study~2 to sketch what innovative, plausible interventions could look like in practice.

\subsection{\revision{Synthesis of Findings}}

\subsubsection{Optimizing the environment}

HCI has made strong progress on environmental control for focused work. Systems reduce external interruptions and visual/auditory clutter, aligning with flow’s need for sustained attention. FlowLight reduces co-worker interruptions via ambient signaling \cite{zugerReducingInterruptionsWork2017a}; cross-device and site blockers curb self-interruptions at opportune moments \cite{kimTechnologySupportedBehavior2017,tsengOvercomingDistractionsTransitions2019}; and VR workspaces dampen open-office noise and visual distractions \cite{ruvimovaTransportMeAway2020,cuberExaminingUseVR2024}. In our landscape, these map to \emph{no distractions}, \emph{quietness}, \emph{being alone}, and elements of \emph{prepared workspace}—interventions that are widely endorsed (Cluster~C) and consistent with flow theory’s preconditions for deepening concentration and clear feedback loops.

At the same time, many workers rely on \emph{background stimulus} (e.g., music, ambient sound) not merely as a shield but to regulate energy and mood. Evidence for background input in cognitive work is mixed and person-specific, which mirrors the polarization we observed (Cluster~B). Current systems rarely modulate such stimulus dynamically as a function of task and state. This leaves space for environment support that goes beyond shielding toward \emph{calibration}: verifying a \emph{prepared workspace} before deep work, sensing ambient load and recommending relocation when necessary, and adjusting background soundscapes over time—for instance, starting suitable audio to lift energy and then automatically fading or simplifying when interaction traces or physiologic signals suggest deepening focus. Participants in Study~2 explicitly imagined \emph{smart workspace checklists}, environment presets for lighting/temperature/seating comfort, and \emph{workplace-selection} aids that detect and recommend quieter spots. Together, these ideas retain Cluster~C defaults while opening up adaptive, person-sensitive support for the more controversial interventions situated in Cluster~B.

\subsubsection{Planning and organization}

A second strand of existing work focuses on this category: Pomodoro-style locking, opportune break/return cues, and transition aids operationalize \emph{single-tasking}, \emph{time boxing}, and \emph{regular breaks} \cite{kimTechnologySupportedBehavior2017,tsengOvercomingDistractionsTransitions2019}. These align with evidence that coherent, less fragmented days foster flow \cite{pluut2024and} and that time is often required for absorption to emerge \cite{peiferAdvancesFlowResearch2021a}. Beyond those approaches, in our data, \emph{time pressure} appears frequently in lived practice yet polarizes in endorsement (Cluster~B). Static, untuned pressure can harm engagement; conversely, bounded, well-timed urgency can help some people and tasks initiate and sustain focus. Despite this, tools that let urgency be \emph{dosed} and \emph{timed} remain rare.

Study~2 participants’ visions point to a path forward: assistants that \emph{auto-schedule} coherent work blocks, \emph{break down} tasks into proximal steps, and apply \emph{gentle, bounded time pressure} when it helps. Here, \emph{urgency becomes a parameter}—calibrated to task type, signals of progress, and personal preference. For example, soft countdowns can nudge initiation, pressure can recede once flow indicators appear (to avoid over-arousal), and blocks can be rescheduled when fragmentation risk rises. Participants also described \emph{voice-to-text capture} for frictionless cognitive offloading, automatic \emph{progress charts}, and \emph{break reminders}. The novelty is not the widgets themselves but \emph{intelligent timing and minimal friction}: capturing ideas at flow-entry edges, showing only proximal progress during deep work, and deferring analytics to cool-down phases so that goal clarity and feedback are maintained without derailing attention.

\subsubsection{Shaping the task}

Clear goals, proximal feedback, and challenge–skill alignment are the classic prerequisites from flow theory \cite{csikszentmihalyiFlowPsychologyOptimal1990,peiferAdvancesFlowResearch2021a}. Existing systems operationalize these through difficulty calibration and structured guidance \cite{aferganDynamicDifficultyUsing2014,ewingEvaluationAdaptiveGame2016,beyrodtSociallyInteractiveAgents2023}, through progress visibility (\emph{progress tracking}), and through metacognitive prompts (e.g., mental contrasting vs.\ goal setting) that help align work with skills \cite{bartholomeyczikFosteringFlowExperiences2023a}. In our mapping, these interventions sit firmly in Cluster~C: broadly endorsed, generalizable, and theoretically well anchored.

Our repertoire, however, pushes \emph{meaningful task} alongside \emph{clear goals}. Flow research often treats meaning as an outcome of the flow state rather than a condition for it. Treating meaning (or task importance) as a designable feature could be very interesting as it has been shown to moderate the relationship between skill--demand balance and flow \cite{engeserFlowPerformanceModerators2008}.  Study~2 participants’ ideas gesture toward this direction indirectly: they asked for tools that decompose work into tangible next steps, make near-term progress visible, and keep attention on the “why” without adding overhead. In practice, this could look like \emph{lightweight value prompts} at planning time, similar to the work of Bartholomeyczik et al. \cite{bartholomeyczikUsingMentalContrasting2024a}, and progress cues that emphasize outcomes over raw activity. \revision{Other ideas suggest that systems could contribute here by automating routine, low-value tasks so that workers can invest their limited attention in the more interesting and meaningful parts of their workday.}

\subsubsection{Mental and physical readiness}

Compared to environment and task shaping, readiness has seen fewer HCI applications despite its centrality in lived practice. Our repertoire highlights \emph{physical activity}, \emph{stimulants/caffeine}, \emph{mindfulness and meditation}, \emph{daylight}, and \emph{nourishment} as practical levers people already use. Such factors have found implementation often in general wellness apps \cite{prasetyoFoodbotGoalOrientedJustinTime2020, kumarExploringUseLarge2023, goddardSystematicReviewFlow2023a, cuiCoDrinkExploringSocial2021}, yet rarely in direct relation with work, let alone with fostering flow as a goal. This leaves opportunity for \emph{embedded} readiness that coordinates with task structure. Participants in Study~2 wished for tools that couple work rhythm with \emph{micro-movement} (\revision{e.g. suggest short active breaks at natural seams, with minimal flow disruption, for example when task switching is already needed}), \emph{daylight prompts} (e.g., mid-morning exposure) and \emph{caffeine timing} (earlier in the day for those who benefit). Because many readiness levers are polarizing (Cluster~B), opt-in personalization and conservative defaults are essential. Elevating readiness from a side-channel into a first-class, low-friction companion can complement Cluster~C fundamentals and provide additional entry points into flow.

\subsection{Future Work}
Looking ahead, future work can build on this foundation in several directions. First, we see opportunities for further implementation of flow interventions in workplace support systems. Our repertoire and design directions offer a menu of candidates that can be developed either as single-intervention tools (e.g., systems for temporal guidance or background calibration) or in multi-intervention solutions that combine several strategies into one coherent whole. Ultimately, this could lead to holistic systems that integrate sensing, orchestration, and personalization to actively support workers across multiple facets of their flow experience.

Second, our work opens the door to flow intervention discovery systems. Rather than only delivering interventions, systems could support workers in exploring which strategies suit them best. This could take the form of recommender systems that suggest interventions based on patterns across users, or lightweight “naïve systems” that present the repertoire directly and enable structured documentation of personal experiences. Such discovery tools would respect the individuality of flow while supporting the self-experimentation that workers already seem to engage in.

Beyond immediate applications, longitudinal and within-person studies are needed to examine how interventions unfold over time, how they interact with individual traits and contexts, and how adaptive personalization can be implemented without undue complexity or disruption. Further, broadening the focus beyond knowledge work—to creative, collaborative, or industrial domains—will test the transferability of our findings and expand the design space.

By combining lived experience with adaptive technology, future systems can progress from merely protecting workers against interruptions to actively enabling them to thrive in flow.

\subsection{\revision{Contributions}}
This paper advances the study and design of digital flow interventions in knowledge work along empirical, theoretical, and design dimensions. We provide empirical insights into the diverse, situated ways in which digital workers actively foster flow, including strategies overlooked in prior research. Study 1 surfaces a diverse repertoire of 38 lived interventions—organized into four actionable themes spanning environment, planning/organization, task shaping, and personal readiness—derived through reflexive thematic analysis of workers’ own practices. Study 2 validates and extends this repertoire at scale, quantifies helpful vs. unhelpful perceptions, and reveals polarization for specific strategies (e.g., background stimulus, time pressure). We further \emph{structure} this repertoire by identifying which interventions are broadly endorsed (“safe bets”), which are \emph{controversial} and therefore highly dependent on the individual and situation, and which are generally disendorsed yet show \emph{niche} value for some users
From this mapping, and by triangulating related work with participants’ envisioned support, we derive design opportunities that go beyond the field’s current emphasis on shielding and static scaffolds. 
Finally, for individual digital workers, our collection serves as a practical starting point to explore and tailor interventions for optimizing everyday flow.

\subsection{\revision{Limitations}}
Our findings should be interpreted in light of several limitations. First, the work is domain-specific: we focus on digital knowledge work. While many lived interventions may transfer to creative, industrial, or highly collaborative settings, their mechanisms and feasibility can differ substantially across domains; external validity beyond desk-based knowledge work remains to be established.

Second, the catalog we derived is not exhaustive. Open-ended elicitation increases breadth, but unavoidably misses strategies some people use (false negatives) and may surface strategies that appear helpful without being causally effective (false positives), especially for interventions only few participants have tried. Relatedly, Study 2 relies on the assumption that people can judge which strategies actually help them. This is partly supported by prior work on flow metacognition and by the theory’s interview-driven origins \cite{wilsonFlowMetacognitionsQuestionnaire2016a, csikszentmihalyiFlowPsychologyOptimal1990}, but self-reports can be shaped by “folk wisdom” or urban myths. Larger samples can average out idiosyncrasies yet will not remove systematic misconceptions. We note, however, that most high-frequency codes align with established literature and theorized preconditions of flow, lending face validity to participants’ judgments.

Third, Study 2 is cross-sectional and aggregate. Participants reported whether a strategy generally helps or does not, but we did not capture within-person, episode-level fluctuations or task contingencies. This limits our ability to explain polarization we observed (e.g., for time pressure or background stimulus): effects likely depend on person factors (e.g., traits, skills), task characteristics (e.g., difficulty, boredom, creative vs. routine work), and situational context (e.g., noise, timing). Although we probed the aim of support (enter, boost, maintain), this only partially addresses context-dependence. As one illustration, Study 1 responses often framed background music as helpful specifically for boring or repetitive tasks; a participant in Study 2 might still answer “does not help” when judging across all their work, masking situational benefits. Longitudinal, within-person designs (e.g., ESM) combined with unobtrusive sensing would be better suited to model these contingencies one strategy at a time.

Fourth, distinguishing flow from adjacent constructs is inherently challenging. Several strategies may improve focus, reduce stress, or enhance well-being without reliably producing flow. We nonetheless consider a flow-centric lens valuable: flow provides a clearly defined, measurable target that integrates productivity with positive experience rather than optimizing one at the expense of the other.

Fifth, measurement and interpretation limits apply. Despite careful wording and pilot refinement of the short descriptions used in Study 2, misunderstandings of specific codes are possible. We mitigated this by providing concise, concrete descriptions, but residual misinterpretation cannot be ruled out.

Finally, sampling considerations constrain generalizability. Although we balanced recruitment by sex recorded on legal documents, we did not query self-identified gender, so we cannot quantify the gender diversity present in our sample. Likewise, we did not account for neurodiversity in screening. Conditions such as ADHD can meaningfully shape how individuals regulate attention and respond to flow-supportive practices; future work should purposively sample and analyze neurodiverse populations.

\section{Conclusion}
We examined how digital knowledge workers foster flow in everyday practice and how these lived interventions can inspire future flow-adaptive systems. Across two studies, we surfaced and validated a repertoire of 38 strategies, showing that some are broadly endorsed, while others are highly individual or contested. This highlights that flow support is not one-size-fits-all but requires adaptive personalization.

Our contribution reframes flow support in HCI as an adaptation problem: while broadly endorsed strategies (Cluster C) provide reliable entry points, going beyond these classics requires personalization to individual needs and tools that help users discover what works for them. Beyond distraction control and skill-demand balancing, we point to overlooked pathways such as readiness management, background calibration, and metacognitive guidance. 

In sum, our findings extend the understanding of flow support by revealing a structured landscape of interventions and pointing to design opportunities that move beyond static support. Consensus strategies offer robust defaults, but the real potential for digital flow interventions lies in personalization and discovery: helping people identify, adapt, and refine the strategies that fit their person and contexts.

\bibliography{hacking_flow}

\end{document}